\documentclass[10pt, conference, compsocconf]{IEEEtran}
\usepackage{mathptmx}
\usepackage[table]{xcolor}

\usepackage{fancyhdr}
\usepackage[normalem]{ulem}
\usepackage[hyphens]{url}
\usepackage[sort,nocompress]{cite}
\usepackage[final]{microtype}
\usepackage[keeplastbox]{flushend}

\usepackage{cite}
\usepackage{amsmath,amssymb,amsfonts}
\usepackage{graphicx}
\usepackage{subcaption}
\usepackage{textcomp}
\usepackage{xcolor}
\usepackage[hyphens]{url}

\usepackage{adjustbox}
\usepackage{mathtools}
\usepackage[noend,ruled,vlined,linesnumbered]{algorithm2e}
\usepackage{multicol, multirow, blindtext}
\usepackage{booktabs}
\usepackage{arydshln}
\usepackage[most]{tcolorbox}
\let\labelindent\relax
\usepackage{enumitem}
\usepackage{soul}

\usepackage[bookmarks=true,breaklinks=true,letterpaper=true,colorlinks,citecolor=blue,linkcolor=blue,urlcolor=blue]{hyperref}

\tcbset{
    frame code={}
    center title,
    left=0pt,
    right=0pt,
    top=0pt,
    bottom=0pt,
    colback=yellow!70,
    colframe=white,
    width=\dimexpr 0.5\textwidth\relax,
    enlarge left by=0mm,
    boxsep=5pt,
    arc=0pt,outer arc=0pt,
    }

\DeclarePairedDelimiter\ceil{\lceil}{\rceil}
\DeclarePairedDelimiter\floor{\lfloor}{\rfloor}

\newcommand{\OUR} {\texttt{MUSE}}
\newcommand{\OURShort} {\texttt{MUSE}}
\newcommand{\OURShortnf} {MUSE}
\newcommand{\OURnf} {MUSE}
\newcommand{\OURFull} {\texttt{MUSE ECC}}
\newcommand{\GM} {\texttt{m}}

\newcommand{\MOD} {\;\mathrm{mod}\;}

\newcommand{\MDDRold} {144,132}
\newcommand{\MDDRoldMult} {4065}
\newcommand{\MDDRoldMSED} {86.71}
\newcommand{\MDDRoldMinSDC} {144,128}

\newcommand{\MDDRoldMinSDCMSED} {99.17}
\newcommand{\MDDRnew} {80,69}
\newcommand{\MDDRnewMult} {2005}
\newcommand{\MDDRnewMSED} {85.03}

\newcommand{\MDDRnewU} {80,67}
\newcommand{\MDDRnewUMult} {5621}

\newcommand{\MDDRnewH} {80,70}
\newcommand{\MDDRnewHMult} {821}

\newcommand\fakesec[1]{\vspace*{2pt} \noindent \hskip .01in \textbf{#1}}
\newcommand{\OTZ}{1$\rightarrow$0}
\newcommand{\ZTO}{0$\rightarrow$1}

\newcommand{\hlc}[2][yellow]{{%
    \colorlet{foo}{#1}%
    \sethlcolor{foo}\hl{#2}}%
}

\def\BibTeX{{\rm B\kern-.05em{\sc i\kern-.025em b}\kern-.08em
    T\kern-.1667em\lower.7ex\hbox{E}\kern-.125emX}}
    
\title{Revisiting Residue Codes for Modern Memories\\
\small \normalfont The latest and authoritative version of the document is published here: \href{https://ieeexplore.ieee.org/document/9923862}{https://ieeexplore.ieee.org/document/9923862}} 

\begin{document}
\bstctlcite{IEEEexample:BSTcontrol}

\author{\IEEEauthorblockN{Evgeny Manzhosov,
Adam Hastings,
Meghna Pancholi, 
Ryan Piersma,
Mohamed Tarek Ibn Ziad,
and Simha Sethumadhavan}
\IEEEauthorblockA{Department of Computer Science\\
Columbia University,
New York, New York, USA\\ 
Email: \{evgeny, hastings, meghna, ryan.piersma, mtarek, simha\}@cs.columbia.edu}
}

\maketitle

\begin{abstract}
Residue codes have been traditionally used for compute error correction rather
than storage error correction.  In this paper, we use these codes for storage
error correction with surprising results. We find that adapting residue codes to
modern memory systems offers a level of error correction comparable to
traditional schemes such as Reed-Solomon with fewer bits of storage.  For
instance, our adaptation of residue code -- \OURFull{} -- can offer ChipKill
protection using approximately 30\% fewer bits.  We show that the storage gains
can be used to hold metadata needed for emerging security functionality such as
memory tagging or to provide better detection capabilities against Rowhammer
attacks.  Our evaluation shows that memory tagging in a \OURShort{}-enabled
system shows a 12\% reduction in memory bandwidth utilization while providing
the same level of error correction as a traditional ECC baseline without a
noticeable loss of performance.  Thus, our work demonstrates a new, flexible
primitive for co-designing reliability with security and performance.

\end{abstract}

\begin{IEEEkeywords}
    error correcting codes; memory tagging; metadata; rowhammer
\end{IEEEkeywords}

\section{Introduction}
Error Correcting Codes (ECCs) are a standard technique for improving system
reliability and are widely used today. ECCs improve reliability by encoding data
in a redundant format that uses additional bits of information to identify and
correct data bits that change while in storage or transit. To minimize storage
and transmission overheads, codes that use fewer redundancy bits for a desired
level of reliability and are easy to implement have gained widespread use. 

Residue codes are a specific type of ECC used for detecting and correcting
errors that happen during computation\cite{STAR,arch_ser} and were developed in
the 1960-70s\cite{henderson_residue_1961,chien_linear_1964,stein1964}. Unlike
storage-oriented codes like Reed-Solomon codes\cite{reed1960polynomial}, residue
codes have a nice property that the error correction information can be computed
in parallel with computational operations: say $e$ is the ECC function, and $f$
is the computational function, then for residue codes, $e(f(x, y)) = f(e(x),
e(y))$ for some common functions $f$. In contrast, for storage-oriented codes,
$e(f(x,y))$ can be calculated only after $f(x,y)$ is computed\cite{arch_ser}.

In this work, we present \OUR{}, our adaptation of the residue codes that
provides a level of protection similar to storage-oriented codes but \emph {with
fewer bits of storage}. These saved bits can be used for storing metadata in a
manner that also protects the metadata itself. As compute and memory are
merging\cite{lee2021hardwarepim}, and security and reliability place more
demands on memory\cite{Califorms,zero_pointer,MTE}, codes, such as \OUR{}, not
only offer a path for better integration of both the compute and memory
components but also enable techniques that need metadata to improve security or
performance.

Residue codes are easy to explain. To protect the data, we multiply it
with a deliberately chosen multiplier before storing it in memory. When the data
is retrieved from memory, we divide the value by the same chosen multiplier. If
there are no errors between the storage and retrieval, we should get a remainder
of zero. If some bits are flipped, we use the remainder to locate the bits that
need to be corrected\cite{henderson_residue_1961}.

To use the remainders to correct a flipped bit, we need each remainder to have a
one-to-one mapping with the error that occurred in that bit. Hence, we enumerate
all possible error patterns, compute their values, and search for a multiplier
that leaves a unique remainder for each error. A good multiplier is the smallest
integer number that satisfies the unique remainder property, as a smaller number
requires fewer redundancy bits.

One challenge with adopting the above code to memory is that memory has to
accommodate failure models where multiple bits can fail. In this case, the
number of remainders grows exponentially with the number of errors, and the
multiplier values become too large for practical
use\cite{mandelbaum_arithmetic_1967, barrows1966new}. To overcome this problem,
we invent a new optimization technique we call \emph {shuffling}. As the name
suggests, shuffling changes the bit positions of the data before it is written
to memory, which changes the distribution of values representing the errors and
their remainders. With shuffling, we can evaluate the same multiplier multiple
times, increasing the chance of finding a multiplier that gives unique remainders.

Another challenge with using this code for memory systems (or even compute
systems) is that a na\"{\i}ve implementation will be prohibitive in terms of
performance as it requires expensive multiplication and division operations on
the critical path delaying memory reads and writes. For instance, using division
can take up to 70 cycles~\cite{fog2020instruction}. We show that we can avoid
these costs using two optimizations: First, the \OURShort{} multiplier is fixed
and known at design time which allows us to avoid the cost of general multipliers
and dividers. We find that specialization of these circuits can reduce the
latencies to about three cycles. Second, we use a systematic encoding of the
codes that mitigates the critical path latency of these workloads.

With these optimizations, we adopt residue codes to modern memory systems and
compare them to the ChipKill scheme based on the Reed-Solomon codes with the
redundancy of commercial schemes~\cite{devicesbios,power9ras} and demonstrate
the following benefits:

\begin{itemize}[leftmargin=*]
\item {\bf \OURShort{} uses fewer bits of storage than comparable ChipKill
schemes.} \OURShort{} corrects multi-bit errors confined to a single DRAM chip
on a DIMM, allowing the system to withstand a permanent memory chip failure with
at least four fewer redundancy bits than ChipKill codes used in enterprise
systems\cite{devicesbios,power9ras}. We show how these saved bits can be used to
implement ARM Memory Tagging Extensions\cite{MTE}, saving up to 3\% of DRAM
power consumption, and how the bit savings can be used to store cryptographic
hashes that reduce the probability of successful Rowhammer\cite{safeguard}
attack to $2^{-40}$.

\item {\bf \OURShort{} is a single code that can guarantee the correctness of
both data in storage and during computation.} \OURShort{} is a promising fit for
Processing In-Memory -- a new emerging technology where computation happens near
the data in memory\cite{lee2021hardwarepim}. We show that in the settings of
PIM-enabled HBM2,~\OUR{} could use $2.6\times$ fewer redundancy bits than the
provisions specified in the standard while protecting the data during both the
storage and the computation without the need for multiple reliability schemes.

\item {\bf \OURFull{} is more flexible.} \OURShort{}'s codewords division
between the redundancy and data is tunable at single-bit granularity for any
symbol size. This flexible division allows to design a code with a specific
target for the spare storage. We show that for multiple configurations of the
codeword, data, and redundancy lengths, \OURShort{} provides ChipKill
guarantees, while Reed-Solomon does not due to its fixed two-symbol redundancy.

\item {\bf \OURShort{} offers customization of codes to fit error models.} We
show how we can cover two classes of errors simultaneously: (1) single-bit
errors and (2) asymmetrical multi-bit errors (i.e., errors due to lack of
refresh) confined to a single DRAM chip on a DIMM. We showcase \OURShort{} code
for this model, which uses 13 check bits leaving three bits for other uses.
\end{itemize} 

The rest of the paper is organized as follows: Section~\ref{sec:bg} provides an
overview of residue codes, Section \ref{sec:muse} describes the construction of
\OUR{}, Section \ref{sec:pc} presents some practical codes, Section \ref{sec:hw}
details the microarchitecture, Section \ref{sec:ben} presents the uses cases,
and in Section \ref{sec:eval} we evaluate the overheads of \OUR{} and compare to
Reed-Solomon codes. We discuss related work in Section \ref{sec:rel} and conclude
with Section \ref{sec:con}.

\section{Background}\label{sec:bg}
In this section, we briefly cover the formulation of the residue codes. We
discuss residue code construction, including its systematic formulation, and the mapping
of errors to remainders. 

\fakesec{Code Construction}
The following equations describe how $data$ is encoded into a
codeword, how a codeword is decoded back into $data$, and how the errors
are detected and corrected:
\begin{equation}\label{eq:mapEq}
\begin{split}
codeword = \GM{} \times data
\end{split}
\end{equation}
\begin{equation} \label{eq:det}
\begin{split}
remainder = codeword\MOD{}\GM{}
\end{split}
\end{equation}
\begin{equation} \label{eq:dec}
 data=
 \begin{cases}
  codeword/\GM{} &remainder=0 \\
  \frac{codeword-f_{error}(remainder)}{\GM{}} &\text{otherwise}
 \end{cases}
\end{equation}
The $data$ is encoded to the $codeword$ by multiplying $data$ by a whole
number~\GM{} (Eq.~\ref{eq:mapEq}). Upon reading the $codeword$, we compute the
remainder of the $codeword$ divided by~\GM{} (Eq.~\ref{eq:det}). If the
remainder is zero, we can recover the original data by simply dividing the
codeword by~\GM{} (error-free case); otherwise we subtract the $f_{error}$ derived
from remainder $r$ to correct the $codeword$ and then recover the data by
dividing by \GM{} (Eq.~\ref{eq:dec}). For this scheme to work, we pick $m$ so
that each remainder corresponds to one error only, allowing for unambiguous
error correction.

The construction above presents a serious problem---the original data 
is available only after division. This can introduce a steep penalty because
of the latency of the division. To avoid this steep penalty for the common
case of no errors, we can use a simple math trick developed by Chien in 1964
\cite{chien_linear_1964}. Chien proposed to separate the $data$ from the
redundancy $r$ by shifting the $data$ in the $codeword$ left by $r$ bits such
that the remaining $r$ bits can be used to store a special value $X$, which is
chosen in such a way so that the $codeword\MOD{}\GM{}=0$ as shown in
Eq.~\ref{eq:encrc}:
\begin{equation} \label{eq:encrc}
\begin{split}
codeword = data << r + X\\
X =\GM{} - (data << r)\MOD{}\GM{}
\end{split}
\end{equation}
This way, the data in the codeword are easily separated from the redundancy,
i.e., $X$, and no integer division is required to recover the data. We will use
this formulation as a basis for \OUR{} so that the data and the error correction
bits can be stored separately (also known as ``systematic'' encoding).

\fakesec{Mapping Remainders to Bit Positions} \label{sec:1b} Next we will
describe how single-bit errors are handled. When an error flips bits, it
transforms the codeword into another codeword. The difference between the
original and erroneous codeword is what we refer to as the \emph {error value}.
However, unlike in traditional codes where single bit flip corresponds to a
single error value, in residue codes, each bit flip has two error values: one
for \ZTO{} bit flips and the other for \OTZ{} bit flips.

As an illustration, let us look into an example. Say, the integer value $243$ is
the codeword. Its binary representation is: $1111\:00\textbf{1}1_2$. Now assume
that bit $b_1$ (in \textbf{bold}) is flipped and codeword value becomes
$1111\:00\textbf{0}1_2$ or $241$, leading to the \emph {error value} of $-2$.
Now let us assume that the codeword is $972$ or $0011\:1100\:11\textbf{0}0_2$.
If the same bit $b_1$ was flipped \ZTO{}, corrupting the codeword to be
$0011\:1100\:11\textbf{1}0_2 = 974$, the \emph {error value} would be $+2$.
Hence, any single bit error may have two distinct error values, and the value
itself depends on the direction of the bit flip. To identify the corrupted bit
and correct its value, every bit in the word requires two distinct remainders.
Note that both of those error values are integers powers-of-two: $-2=-1\times
2^1$ and $2=1\times 2^1$. Hence, in general, error at bit $b_i$ has error values
of $E_{i}=\pm2^i$, where a positive value is for bit error \ZTO{}, and a
negative value is for the error \OTZ{}.

\section{\OURnf{}}\label{sec:muse}
The material presented in the previous section was developed back in the 1960s
and 70s in the context of computers that featured arithmetic
units\footnote{Caches were invented in 1965\cite{wilkes1965slave} and DRAM in
1967\cite{dram_patent}}. While the math works well for memories too, directly
applying the ideas to memories is not straightforward. For instance, just being
able to support single-bit errors is not very useful for memory errors; we need
multi-bit error support as these errors are common enough to warrant support for
a large class of systems. However, introducing multi-bit error support
(naively) increases the storage requirements of the code
\cite{mandelbaum_arithmetic_1967, barrows1966new}, making it less competitive
with modern codes. In this section, we build upon the construction from the
previous section and extend residue codes to fit modern memory systems.

\subsection{Mapping Remainders to Symbol Errors}\label{sec:symb} We start by
extending the residue codes to protect against multi-bit errors confined to a
single $s$-bit-wide DRAM device. We are interested in this class of errors
because codewords are striped across multiple DRAM chips on a DIMM, and if one
chip fails, the corresponding part of the codeword will exhibit an $s$-bit
error. Hereon, we define a symbol as a group of $s$ bits written to a single
DRAM device. Each $s$-bit symbol may have $2\times(2^s-1)$ possible errors---
same as the number of ways for one $s$-bit number to become a different $s$-bit
number. For example, for 4-bit symbols, a 4-bit value can be corrupted via 30
different errors, i.e., to become a value other than the original. Some of these
error values are positive while others are negative. If the initial data value was
$0000_2$, it could have only 15 positive errors, while $1100_2$ would have three
positive and twelve negative error values. 

\subsection{Shuffling} \label{sec:shuffle}
\begin{figure*}[t]
	\begin{subfigure}{0.66\textwidth}
		\includegraphics[width=0.99\linewidth]{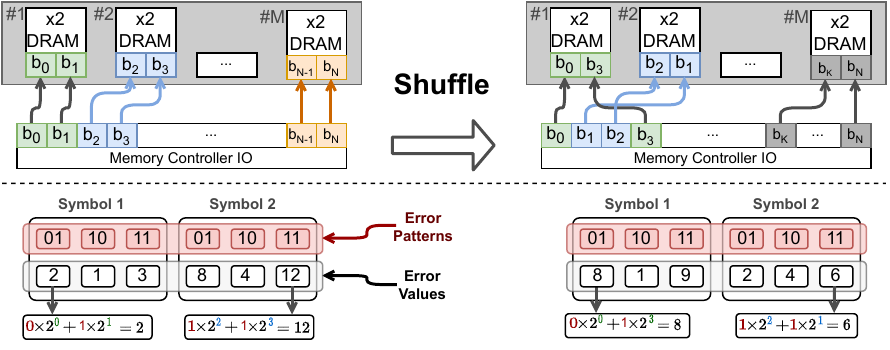}
	 \caption{ }\label{fig:shuf_ab}
	\end{subfigure}
	\begin{subfigure}{0.32\textwidth}
		\includegraphics[width=0.99\linewidth]{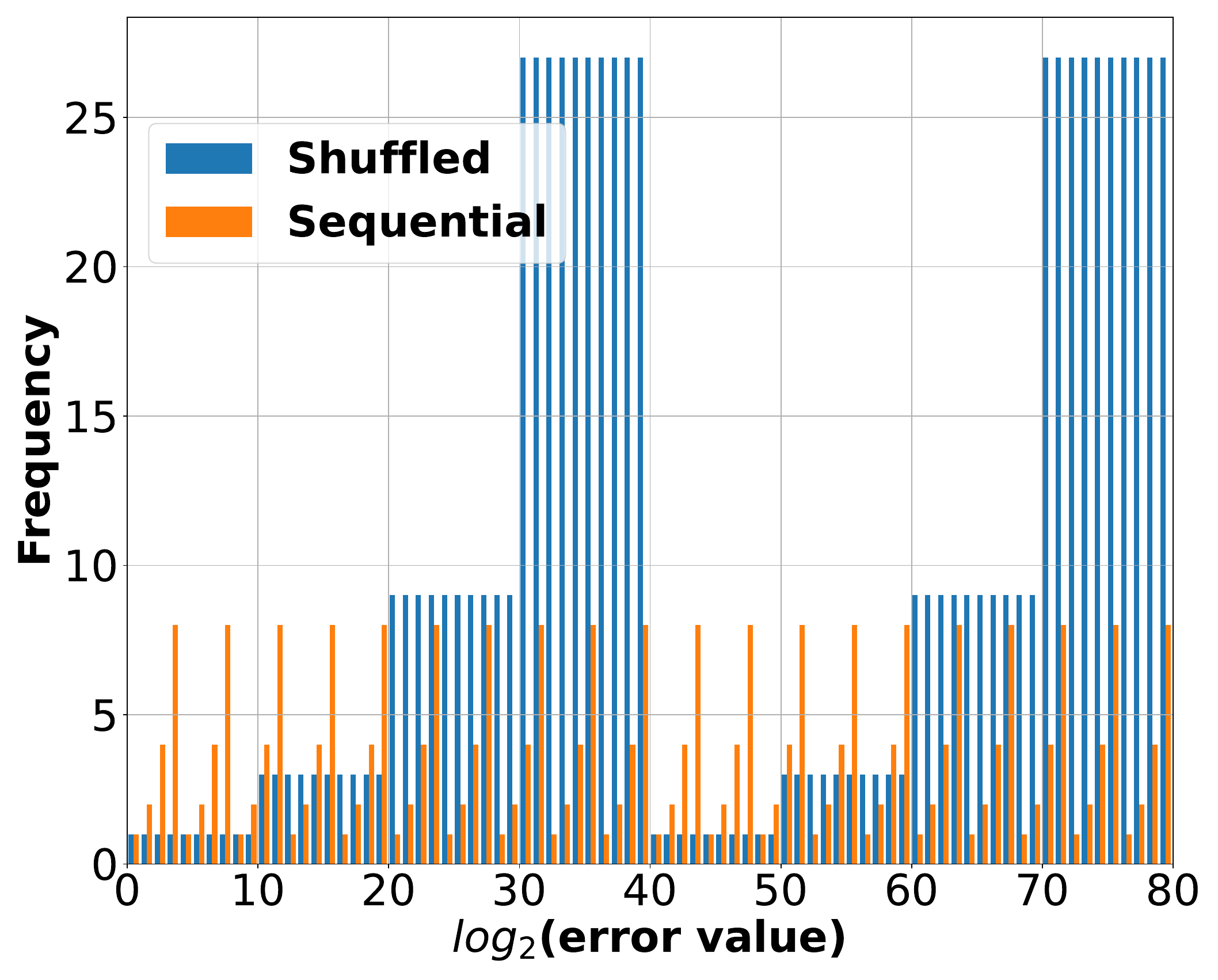}
	 \caption{ }\label{fig:shuf_abc}
	\end{subfigure}
	\caption{ (a) Top: implementation of the shuffling in hardware -- routing
	the signals between the memory controller and DRAM interface in a
	``shuffled'' manner (on the right); bottom: reduction of error values'
	numerical range from [1,12] to [1,9] due to shuffling. After shuffling, DRAM
	$\#1$ has bits (green color) $b_0$ and $b_3$ of the encoded codeword,
	resulting in error value of $8$ instead of $2$ with sequential assignment
	(see equations). Here and thereafter, for convenience, only the positive
	values are shown. (b) Redistribution of error values due to the shuffling
	for \OURShort{}(\MDDRnew{}) code. }\label{fig:shuffling_figs}
\end{figure*}

However, with the symbol error model, the amount of errors in the code increases
significantly. This increase, in turn, leads to a higher chance that multiple
errors will result in identical remainders, making it impossible to differentiate
between them for correction. To overcome this issue, we introduce
\emph{shuffling}---a new technique that reduces the chance of different errors
having the same remainder. 

We explain shuffling and its benefits with a simple example. The top portion of
Figure~\ref{fig:shuffling_figs}(a) shows a toy example where the memory
controller writes the data in a shuffled manner into a DIMM with x2 DRAM
devices. In this example, codeword bits $b_0$ and $b_3$ are written into DRAM
device $\#1$, bits $b_1$ and $b_2$ are placed in device $\#2$, etc. The bottom
portion of the figure shows how shuffling affects the numerical range of the
error values (for convenience, only the positive values are shown): the error
values of the first symbol are 1, 8, or 9, while the error values for the second are 2, 5, or
6. For instance, for the first symbol, the error value of error pattern $01$
changes from $2$ to $8$ (see the equations in the figure). As a result, with
shuffling, the numerical range of error values changes from [1,12] to [1,9].
Thus, shuffling results in a different set of error values, increasing our
chances that a given multiplier can be used to disambiguate the error patterns.

\fakesec{Mapping Remainders to Shuffled Symbol Errors} With shuffling, an $s$-bit
symbol error is transformed from a sequential $s$-bit error into $s$ single-bit
errors. Continuing with our example, failure of DRAM $\#1$ in
Figure~\ref{fig:shuffling_figs}(a) results in corruption of bits $b_0$ and
$b_3$ as opposed to $b_0$ and $b_1$ for the un-shuffled case. Generalizing, the
$s$-bit shuffled symbol error may have one of the $3^{n}-1$ possible error
values. For example, the first symbol of a codeword from
Figure~\ref{fig:shuffling_figs}(a), with bits $b_0$ and $b_1$, has six possible
error values: $\pm1,\pm2$, and $\pm3$. After shuffling, this symbol stores bits
$b_0$ and $b_3$ and has eight possible error values: $\pm1,\pm7,\pm8$, and
$\pm9$. Despite the increase in the number of error values per symbol,
shuffling enables codes that cannot be defined with the traditional residue code
construction~(Section~\ref{sec:pc}).

Figure~\ref{fig:shuffling_figs}(b) shows how shuffling affects the distribution
and the count of code error values of the \OURShort{}(\MDDRnew{}) code. The
code error values are computed for the sequential (orange bars) and shuffled
(blue bars) bit assignments and shown in the histogram. The error values for
the sequential bit assignment are fewer, and their frequency distribution per
bin is not uniform. On the other hand, with shuffling there are more error
values, and their frequency is more uniformly distributed in each bin. However,
as expected, the total number of errors with shuffling is much higher than
without (the area under the blue bars vs. the area under the orange bars).

\fakesec{Construction} The hardware implementation for shuffling is trivial as
it requires only minor changes in signal routing between the DRAM interface IO
and the memory controller IO. On writes, we encode the codeword, shuffle, and
store it to DRAM. When reading the codeword back from memory, we first
un-shuffle it and then decode. 

\subsection{Asymmetrical Errors}\label{subs:uni} Here we consider an
asymmetrical error model where only one type of erroneous transition is
possible: \ZTO{} or \OTZ{}, but not both\cite{asym_dram,asym_dram2}. Where is
this model useful? Using this error correction model is often useful to protect
against data retention errors in DRAM to reduce the power spent on refresh.
Prior research has shown that the majority of DRAM cells can hold the data
longer than the refresh period since DRAM refresh frequency is set to ensure the
reliability of the unreliable minority of memory cells\cite{asym_dram3,raider}.
This conservative approach to DRAM refresh increases the power spent to refresh
the data and, inadvertently, thwarts performance because during the refresh the
memory is unavailable to serve CPU requests. If we had ECC schemes with high
reliability guarantees against those types of errors, they might have saved
power spent on data refresh and improve power and performance.

To illustrate how \OURShort{} can handle asymmetric errors, without loss of
generality, we assume \OTZ{} errors only. Thus, the single-bit error value must
be a negative integer. Similar to symbol errors, the error values of
asymmetrical symbol errors are a combination of negative error values of
individual bit flips within the symbol. Using only the negative half of error
values cuts down the number of required remainders by half, increasing the
chances of finding a one-to-one mapping between errors and remainders.

\subsection{Multiplier Search Procedure}\label{sec:cocon}
Algorithm~\ref{alg:search} is a pseudocode implementation for finding code
multipliers\footnote{See Appendix~\ref{sec:ae} to obtain the C++ implementation.}.
The algorithm inputs are the code length $n$, the symbol size $s$, and
redundancy $r$. The output is the list of multipliers (empty, if none are
found) satisfying the code constraints, i.e., codeword size, symbol size, etc.
We denote a satisfying code as \OURShort{}(n,k).

The procedure starts (line \ref{line:a}) by assigning codeword bits to symbols,
initializing an empty set of valid multipliers $mults$ (line \ref{line:b}), and
precomputing a required number of remainders for the code $R$ (line
\ref{line:c}). For every multiplier $\GM{}$ (line \ref{line:d}), we compute
remainders for error values in the codeword. For each error pattern
(line~\ref{line:f}) in symbol $S_i$ (line \ref{line:e}), we compute its error
values $errVal$ (line~\ref{line:g}) by calling $getErrVals()$
(lines~\ref{line:f1}-~\ref{line:f2}). An empty (line \ref{line:d1}) remainder
set is filled with computed remainders (line \ref{line:i}) of each $errVal$. We
compare remainders' set size to the required number of remainders $R$ (line
\ref{line:r}); if they match, and all the remainders are not zero, we put
multiplier $\GM{}$ into the list of valid code multipliers (line \ref{line:h}).
We repeat this procedure until we exhaustively checked all multipliers in the
redundancy bits $r$. 

\begin{algorithm}[!t]
	\caption{Code Multiplier Search}\label{alg:search}
	\SetKwInOut{Input}{input} \SetKwInOut{Output}{output} \Input{$r$ redundancy
	bits, $s$ byte size, $n$ code length} \Output{List of multipliers $mult$}
	\SetKwFunction{getErrVals}{getErrVals} \BlankLine \emph{$S \leftarrow
	assignBitsToSymbols(s, n)$\label{line:a}}\tcp*[l]{Section \ref{sec:shuffle}}
	\emph{$mults \leftarrow empty()$}\;\label{line:b} \emph{$R \leftarrow
	remaindersNeeded(s, n)$}\label{line:c}

	\For{\textbf{odd}$\:\GM{}\in 2^{r}+1$ \KwTo $2^{r+1}-1$}{\label{line:d}
		\emph{$remSet \leftarrow empty()$}\;\label{line:d1} \For{$S_i\in
		S$}{\label{forins}\label{line:e} \For{$errPattern\in 1 $ \KwTo
		$2^s-1$}{\label{line:f} \For{errVal$\in
		getErrVals(errPattern,S_i)$}{\label{line:g}
		$remSet.insert(errVal\:$mod$\: \GM{})$\;\label{line:i} } } }
		\If{$remSet.size()==R \And 0\not\in remSet$}{\label{line:r}
	
			 $mults.insert(\GM{})$\;\label{line:h} } }
		 \SetKwProg{Fn}{Function}{:}{} \Fn{\getErrVals{$errPattern$,
		 $S_i$}}{\label{line:f1} \emph{$locErrVals \leftarrow empty()$}\;
		 \emph{$binPattern \leftarrow to\_bin(errPattern)$}\; \For{vec$\in
		 genAllVectors(binPattern)$}{ $locErrVals.insert( vec \cdot
		 S_i)$;%
		} \KwRet locErrVals\;\label{line:f2} }
\end{algorithm}

\section{Practical Code Examples}\label{sec:pc}
In this section, we describe~\OURShort{} codes for different error and
system models to showcase the flexibility and applicability of our code in
practical modern contexts. Code parameters, i.e., multipliers and shuffles, are
summarized in Table~\ref{table:museCodes}. We denote by $(n, k)$ a code that
encodes $k$-bit of data into $n$-bit codeword. While both Reed-Solomon and
\OURShort{} can be designed to handle multiple errors, in this section, we focus
only on single-symbol correcting codes as those are often used to guarantee
ChipKill in commercial systems. 

To help classify the codes, we propose the following naming convention based on
the error type covered by code: $PST$, where $P$ is the error constraint form,
$S$ is the size of the error, and $T$ is error type.  We support two types of
errors: $B$---bidirectional, i.e., bit flips in both directions, and $A$---
asymmetrical, i.e.,  bit flips in one direction only, i.e., retention errors in
DRAM.  Both types of $S$-bit errors may be either $S$-bit constrained---marked
by $C$, or unconstrained, i.e., any consecutive $S$-bit pattern, marked by $U$.
For example, $C4B$ code covers any 4-bit bidirectional error in the codeword,
where each $4$-bit error may start at bit $\#0$, bit $\#4$, etc., while $U4B$ is
a $4$-bit errors that may start at any bit position.

This naming approach highlights the flexibility of \OURShort{} design. For
instance, \OURShort{}($\MDDRnewH{}$) covers constrained $4$-bit asymmetrical
errors and any $1$-bit bidirectional error. In this case, we would name this
code $C4A\_U1B$. 

\begin{table}[t]
 \centering
 \caption{Design parameters of~\OURShort{} codes.}
 \label{table:museCodes}
\scalebox{0.990}{
\begingroup
 \begin{tabular}{c c c c}
  \toprule
 \multicolumn{2}{c}{\textbf{Code}} &  \multirow{2}{*}{\textbf{multiplier} \GM{}
 }&  \multirow{2}{*}{\textbf{shuffle}} \\
 \textbf{name} & \textbf{type} & &   \\\hline
 \OURShort{}($\MDDRold{}$) 		& $C4B$ 		& \MDDRoldMult{}  	& None \\
 \OURShort{}($\MDDRnew{}$) 		& $C4B$ 		& \MDDRnewMult{}  	& None \\
 \OURShort{}($\MDDRnewU{}$) 	& $C8A$ 		& \MDDRnewUMult{} &
 Eq.\ref{eq:bitAssign}\\
 \OURShort{}($\MDDRnewH{}$)  	& $C4A\_U1B$ & \MDDRnewHMult{} &
 Eq.\ref{eq:bitAssign2}  \\ 
 \bottomrule
 \multicolumn{4}{l}{\textit{Legend:}}\\
 \multicolumn{2}{l}{\textbf{C} Constrained symbols} &
 \multicolumn{2}{l}{\textbf{A} Asymmetrical errors} \\
 \multicolumn{2}{l}{\textbf{U} Unconstrained symbols} &
 \multicolumn{2}{l}{\textbf{B} Bidirectional errors}\\
\end{tabular}
 \endgroup
 }
\end{table} 

\fakesec{\OURShortnf{}(\MDDRold{}) Single Symbol Correct (SSC) Code} Server and
enterprise machines often require a specific capability known as ``ChipKill''.
Informally, ChipKill allows operation even when one or more DRAM chips on a DIMM
completely fail. While the DDR5 standard has been
finalized\cite{jedec2020jesd796} and devices are being sold, older DDR4 devices
will likely continue to be utilized for at least a few more years. Hence, we
show how \OURShort{} can be used to provide ECC for DDR4 DIMMs. Rather than
designing for \emph {device} failures, we design for handling \emph {symbol}
failures, as symbols are at least as large as DRAM devices and usually a
multiple of the device size. Thus, SSC code guarantees to correct errors
originating in a single device on a DIMM.  

To correct single device failure with x4 devices, we use the
\OURShort{}(\MDDRold{}) code with multiplier $\GM{}=\MDDRoldMult{}$ and 4-bit
symbols. In our scheme, the data bits are striped across two DIMMs with 18 x4
devices each, forming a 144-bit channel\footnote{Those 144-bit channels are
quite common in systems with IBM Power9\cite{power9ras} and Intel
Xeon\cite{supermicro} CPUs}. Under these conditions, \OURFull{} uses
only 12 check bits for 132 bits of data\footnote{We use the term ``data'' to
refer to amount of useful information in the codeword, and not to the native
data sizes.}. In contrast, a traditional Reed-Solomon code will use 16 check
bits for 128 bits of data. 

\fakesec{\OURShortnf{}(\MDDRnew{}) SSC Code} The recently published DDR5
standard\cite{jedec2020jesd796} doubles the number of channels per DIMM,
requiring two 40-bit memory channels per DIMM (32-bit data + 8-bit
parity)\cite{micronDDR5ecc}. As of this writing, there are no commercially
available DDR5 ECC DIMMs, and as such it is not clear how exactly the DIMMs will
be configured. It is possible that they could be made of ten x4 devices per
channel, for a total of 20 devices per DIMM, or five x8 devices per channel, for
a total of ten devices per DIMM.

Like in DDR4, we propose a 4-bit symbol (\MDDRnew{}) code with data striped
across two channels. \OURShort{}(\MDDRnew{}) code encodes 69-bit data into
80-bit codewords, using only 11 redundancy bits. With this code, by using only
64-bits for data, we can correct failure of one device on a DIMM with five bits
to spare; or we can recover two consecutive device-failures with one bit to
spare. In contrast, a traditional Reed-Solomon(80,64) code with x8 symbols has
no spare bits.

\fakesec{\OURShortnf{}(\MDDRnewU{}) Single Device Correct Code} Suppose that
we wish to correct a single device failure, and we do not wish to draw data from
two channels on a DDR5 DIMM. Even in this case, we can design a \OURShort{}
code. Assuming asymmetrical errors, we can design a code with 8-bit symbols for
DDR5 DIMMs. For this code, sequential assignment of bits to symbols yields no
multipliers of 16-bits or less. Thus, we \textit{shuffle} the bits between
symbols. As a result we found a multiplier~$\GM{} = \MDDRnewUMult{}$ for the
following shuffle:

\begin{equation}\label{eq:bitAssign} 
 \small
 \begin{split}
S_{i}^{i\in[0,9]} = [b_i,b_{10+i},b_{20+i},b_{30+i},b_{40+i},b_{50+i},b_{60+i},b_{70+i}]
\end{split}
\end{equation}

This \OURShort{}(\MDDRnewU{}) code encodes 64 bits of data and three bits of
metadata into an 80-bit codeword while correcting a single device failure. To
use this code over a 40-bit channel with 80-bit codewords, we split the
codewords such that every bus transaction carries half of the 8-bit symbol to
memory (for all symbols). A similar approach is used by AMD\cite{devicesbios}.  

\fakesec{\OURShortnf{}(\MDDRnewH{}) Single Device Correct Hybrid Code} To
showcase the flexibility of the construction approach, we design a 4-bit symbol
code that handles two classes of errors: (1) asymmetrical symbol errors, and (2)
bidirectional single-bit errors. We call codes that correct more than one type
of error \emph {Hybrid codes}. As a result, the code encodes 64-bit data and
6-bit metadata into 80-bit codewords and corrects two classes of errors. Code
parameters are ~$\GM{} = \MDDRnewHMult{}$ with the following shuffle:

\begin{equation}\label{eq:bitAssign2} 
 \small
\begin{split} S_{2\times i} =
[b_i,b_{10+i},b_{20+i},b_{30+i}] ,\;i\in[0,9]\\ S_{2\times i + 1}
= [b_{40+i},b_{50+i},b_{60+i},b_{70+i}] ,\;i\in[0,9] \end{split}
\end{equation}

\section{\OURnf{} Microarchitecture}\label{sec:hw}
In this section, we discuss the microarchitectural implementation of~\OURShort{}
codes. We start with the system-level integration, discuss the design of the
encoder and decoder, and conclude with lower-level building blocks and
optimizations for fast multiplication and fast modulo operations that are
required to build efficient decoders and encoders.

\subsection{System Overview}\label{sec:sysover} 
\begin{figure}[t]
\includegraphics[width=0.5\textwidth]{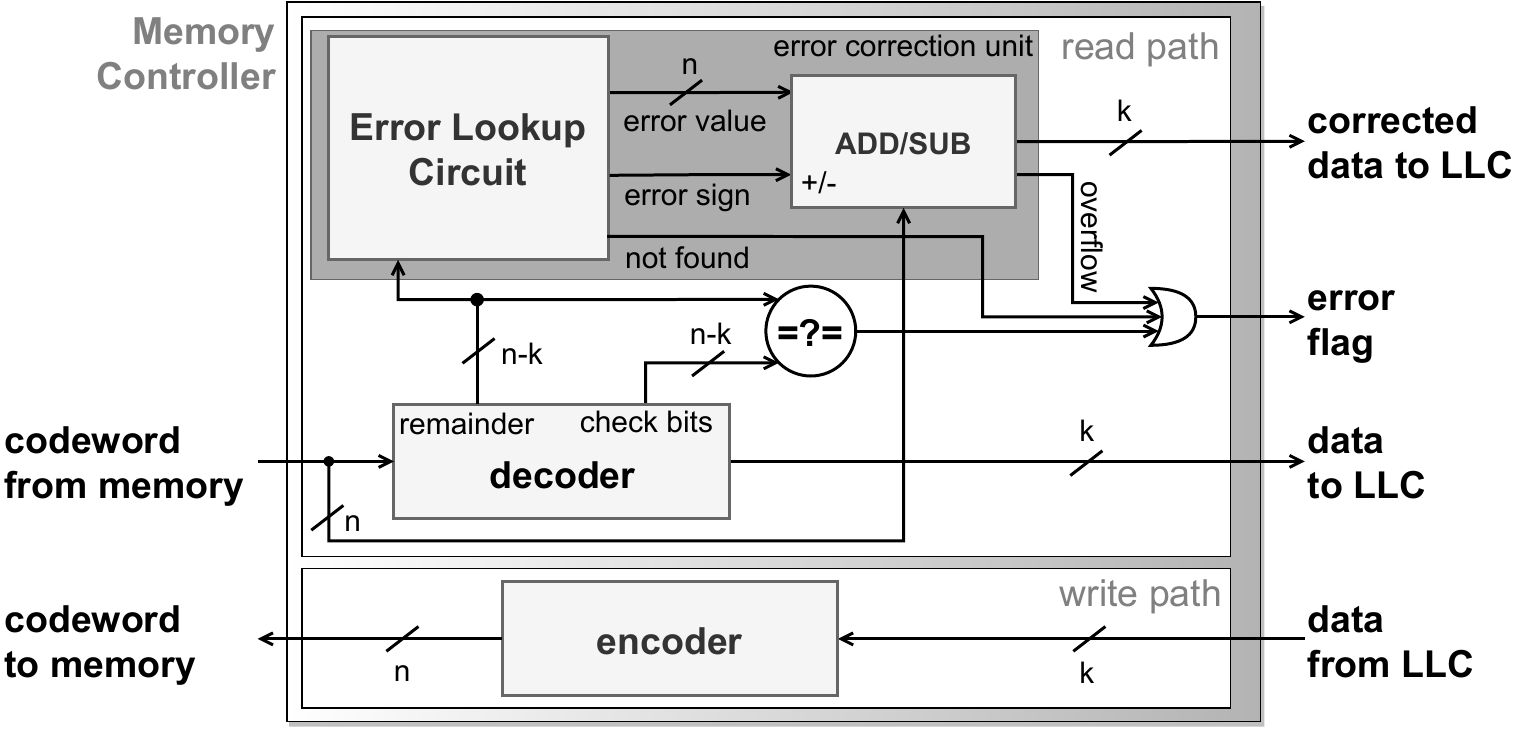}
\caption{Memory Controller with integrated \OUR{}.}
\label{fig:hw_sys}
\end{figure}

Figure~\ref{fig:hw_sys} shows a
high-level overview of \OURShort{} integrated into a system. On the read path,
for every word read from the main memory, the decoder computes the remainder,
which is passed to the Error Lookup Circuit (ELC) to determine whether an error
has occurred. If the ELC provides a value to correct the data, then error
correction is performed, and the Last Level Cache (LLC) will read the corrected
data. On the write path, the data are read from the LLC, encoded, and
transmitted to the memory.

\begin{figure}[t]
\includegraphics[width=0.48\textwidth]{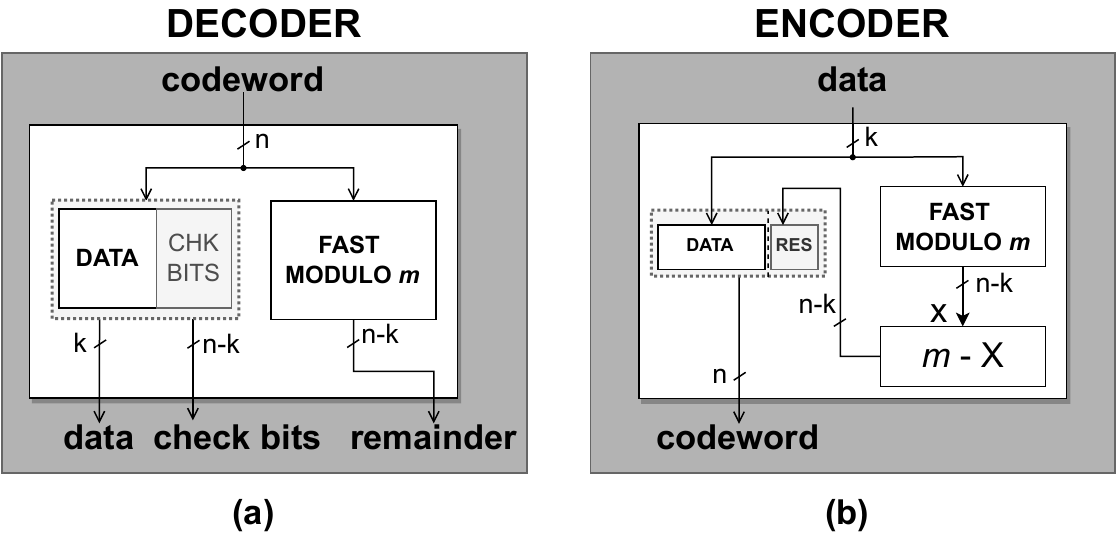}
\caption{ (a) The systematic decoder has zero delay on the critical path due to
the separation of the data and redundancy bits. (b) The systematic encoder first
computes the value of \texttt{X} (Eq.~\ref{eq:encrc}), and then attaches it to
the data making a separable codeword.}
\label{fig:encdecs}
\end{figure}

\fakesec{Decoder} Figure~\ref{fig:encdecs}(a) shows the microarchitecture of the
decoder. The decoder uses a \texttt{fast modulo} (described in
Section~\ref{sec:ariblo}) circuit to compute the remainder, which is used for
error detection and correction. The decoder is systematic, i.e., utilizes the
separability of the data and does not require division like in non-systematic
residue codes.

\fakesec{Encoder} The encoder in Figure~\ref{fig:encdecs}(b) is similar to the
decoder as it computes the residue using \texttt{fast modulo}. However, unlike
the decoder, encoders need to calculate the value of \texttt{X} (see
Eq.~\ref{eq:encrc}) to ensure that~$codeword\MOD{}\GM{}=0$.

\fakesec{Error Correction} Figure~\ref{fig:hw_sys} shows the implementation of
the error correction unit which consists of the Error Lookup Circuit (ELC) and an
adder. Each entry in the ELC contains a remainder, error value, and the sign bit
for the adder. The remainder from the decoder's output is compared against the
remainder field in the ELC entry, and upon a match, the error value is used to
correct the error. The sign bit in the matched entry directs the adder to
subtract or add the error value to the codeword. For example, for
\OURShort{}(\MDDRold{}) code with $\GM{}=\MDDRoldMult{}$, the error correction
is built around ELC with 1080 entries and an adder. Each entry is 157 bits wide,
where the first 12 bits are the remainder value, the next 144 bits are the error
value, and the remaining bit is the adder sign bit. 

\begin{figure}[t]
 \includegraphics[width=0.48\textwidth]{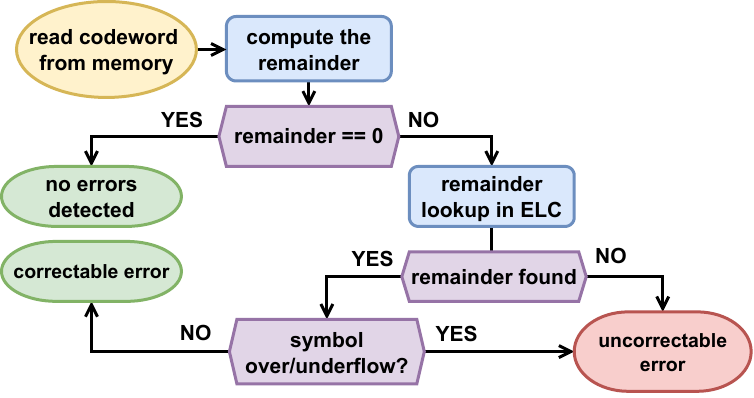}
 \caption{Error detection and error correction decision diagram for the
 \OURFull{}.}
 \label{fig:flowchart}
\end{figure}

\fakesec{Error Detection} There are two ways to detect multi-symbol errors with
\OUR{}: (1) the computed remainder is not present in the ELC, or (2)
symbol-overflow/-underflow during error correction (as shown in
Figure~\ref{fig:flowchart}). The first method utilizes the fact that the code
multiplier is greater than the number of remainders it generates, leading to
some remainder values being unused. Hence, if the ELC does not find the matching
remainder entry, it declares an uncorrectable multi-symbol error. We base
the second detection method on the observation that errors are symbol-confined
by design. Thus, if some miscorrected multi-symbol errors cause flipping bits
beyond the boundary of the corrected symbol, the multi-symbol error is detected.
This method works because the errors are corrected with addition/subtraction,
and in the case of a multi-symbol error, it may cause a ripple of 1s or 0s
beyond the symbol boundary.
 
\subsection{Fast Arithmetic Blocks}\label{sec:ariblo}

 \begin{table}[t]
 \centering
 \caption{Summary of arithmetic operations for codes.}
 \label{table:hw_summary}
 \begin{tabular}{r c }
 \toprule
 
 \textbf{Process} & \textbf{Operations} \\ \hline
 decode & 
  \begin{tabular}{@{}c@{}} $ d = c >> r$ \\ $r = c \mod \GM{}$\end{tabular} \\
 \cdashline{1-2}[1pt/1pt]
 encode & 
  \begin{tabular}{@{}c@{}} $ r = c \mod \GM{}$ \\ $c = d<<r -r$\end{tabular} \\
\bottomrule
\multicolumn{2}{l}{\small $r$ are the redundancy bits, or $r = \ceil*{log_2
\GM{}}$}\\
 \end{tabular}
\end{table} 

We build \OUR{} with three arithmetic operations: integer division,
multiplication, and modulo. Both encoding and the decoding are done by computing
the modulo (see Table \ref{table:hw_summary}).

\fakesec{Division by Constant} The decoder uses modulo for error
detection/correction and relies on a fast dividers and multipliers. However,
even the fastest processors take 13--95 cycles for integer
division~\cite{fog2020instruction}. Two insights help us to substantially reduce
decoder latencies: (1) \emph{a general divider is not required} since we always
divide by a known multiplier of the code~\GM{}. (2) \emph{using multiplication
by the inverse instead of the division}---a known optimization technique in
compilers \cite{gccDivByConst,lemire2019faster,barrett1986implementing}. As a
result, we reduce the problem of designing a fast divider to a task of designing
a fast multiplier by a constant.

\begin{figure}[t]
\includegraphics[width=0.48\textwidth]{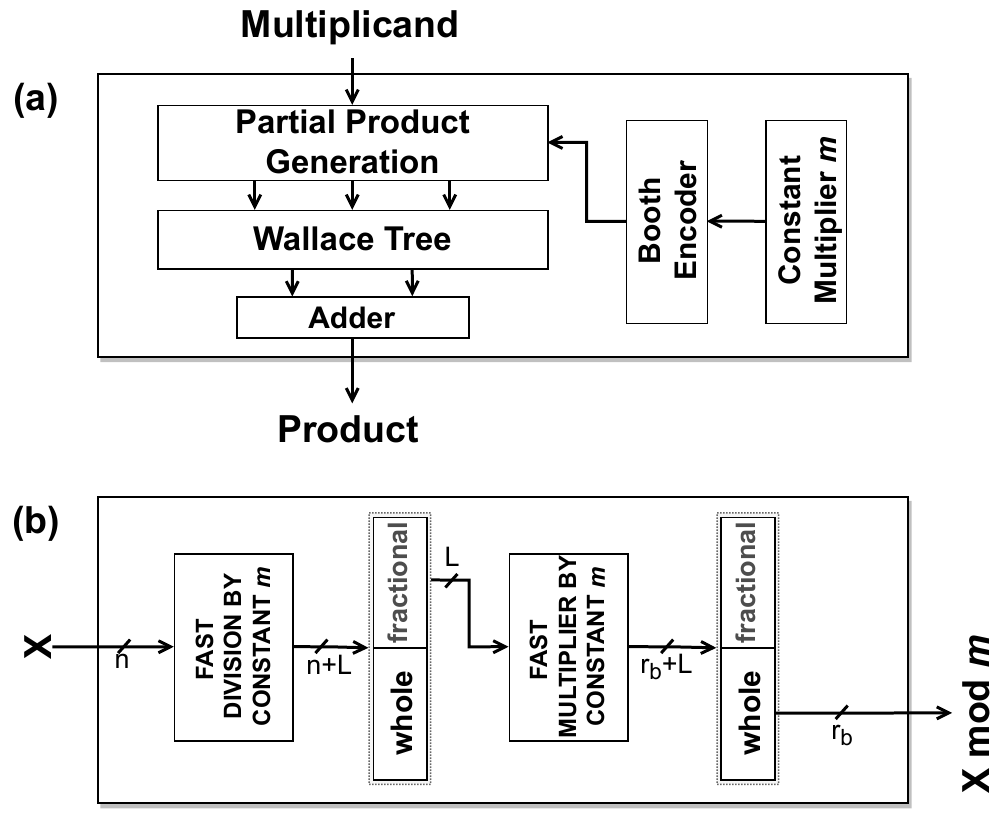}
\caption{(a) The outline of a multiplier with Booth Encoding. Three main
components are Booth Encoding, Wallace Tree, and final adder stage. (b)
Microarchitecture of a circuit for direct remainder computation. Both
multipliers are multipliers from (a), implementing a scheme described by Lemire
\cite{lemire2019faster}.}
\label{fig:arithComponents}
\end{figure}

\fakesec{Multiplication by Constant} Generic integer multiplication of 64-bit
operands is typically done in 3--4 cycles in most modern
CPUs. However, because decoding is on a critical path
and the codewords are at least 80-bit long, we need a much faster multiplier. We
achieve this goal with a custom Wallace Tree\cite{wallaceTree} multiplier based
on Radix-4 Booth Encoders\cite{modBoothRadix4}. Figure
\ref{fig:arithComponents}(a) outlines the multiplier with its three components:
(1) Multiplier encoder with Radix-4 Booth Encoding (BE), which reduces the
number of partial products by half, and makes the tree shallower, (2) Wallace
tree that performs the summation of partial products, and (3) a final adder that
produces the result. We optimized the depth of the Wallace Tree further by
analyzing partial products and removing those always equal to zero from the
multiplier tree, further reducing the latency, hardware, and energy costs of the
multiplier. For example, for the~\OUR{}(\MDDRold{}) code, Booth Encoding of the
multiplier's inverse value has 73 partial products, of which 23 are equal to 0.
By eliminating these, we reduce the depth of the Wallace tree by one level, thus
reducing the latency by three XOR delays. Table~\ref{table:inverses} summarizes
the multipliers, their inverses and shift amounts we used to implement the
codes.

\begin{table}[t]
 \centering
 \caption{Multipliers and their inverses for~\OURShort{} codes.}
 \label{table:inverses}
 \begin{tabular}{l | c}
 \toprule
 $\GM{}$ & \textbf{Inverse Value} \\\hline
$\MDDRoldMult{}$ & $22470812382086453231913973442747278899998963^\ast$\\ 
$\MDDRnewMult{}$ & $77178306688614730355307^\diamond$ \\ 
$\MDDRnewUMult{}$ & 1761878725188230243585305$^{\alpha}$ \\
$\MDDRnewHMult{}$ & 753922070210341214920295$^{\beta}$\\
\bottomrule
\multicolumn{2}{l}{\small shift right by: $\ast$ 156, $\diamond$ 87, $\alpha$
93, and $\beta$ 89 bits} \\
\end{tabular}
\end{table}

\fakesec{Modulo by Constant} The naive approach to compute $c$
\texttt{mod}~\GM{} is as follows:
\begin{equation}\label{eq:naiveResidue}
\begin{split}
r = c - \GM{}\times \floor*{ c / \GM{} } 
\end{split}
\end{equation}
where the division may be a multiplication with the inverse. As a result, the
latency of the modulo operation is the latency of two multiplications and one
subtraction, which can be done in 7 cycles on a modern
CPU\cite{fog2020instruction}. However, the technique developed by Lemire et
al.~\cite{lemire2019faster} allows computing the modulo even faster. The idea is
based on using discarded fractional bits from the division (within $\floor*{\cdot}$
in Eq.\ref{eq:naiveResidue}). When the value represented by these bits is
multiplied by the code multiplier~\GM{}, the upper $r$ bits of the product are equal to the
result of a modulo. Figure~\ref{fig:arithComponents}(b) shows a schematic of a
circuit implementing this scheme with two consecutive multiplications. The
second multiplier is much faster than the first one because it multiplies
by~\GM{} -- a much smaller integer than the inverse of~\GM{}. Therefore, the
resulting latency of a custom modulo circuit is much shorter than the latency of
serialized CPU operations.

\section{\OURShortnf{} Use Cases}\label{sec:ben}
In this section, we discuss co-design opportunities for~\OURShort{} codes.  

\subsection{Exploiting Unused States for Security}
Let us start with an obvious case. Consider a \OURShort{}(\MDDRnew{}) code with
an $11$-bit multiplier $\GM{}=\MDDRnewMult{}$, thus giving us five bits of space
that one can use for metadata. In other words, with an 80-bit codeword, we can
store up to 69 bits of data and still get ChipKill-like functionality, but since
the basic granule of protection is 64 bits (8 bytes) in most existing schemes,
we get to store five additional bits of free storage. This storage can be put to
several uses, for example, the ARM Memory Tagging Extensions (MTE)\cite{MTE} or
Rowhammer detection\cite{rhmr}. We discuss the integration of \OUR{} with ARM
MTE in Section~\ref{case:mte}, and in the next section we discuss Rowhammer
resiliency. 

\fakesec{Rowhammer Detection}
In 2019, Cojocar \emph{et al.}~\cite{eccsploit} showed that it is possible to
bypass the ECC and execute a successful rowhammer-based attack on a system.
Indeed, common ECC schemes are not designed to withstand a high number of bit
flips within a codeword as their detection capabilities are limited by available
ECC storage. Because \OURShort{} is more storage-efficient, it can be used to
mitigate Rowhammer-based attacks (e.g., \cite{yim2016rowhammer}). For example,
we use the salvaged five bits per word provided by \OURShort{}(\MDDRnew{})---
which amounts to 40 bits per cache line---to store a hash code of the cache
line's worth of data. Hence, when the attacker rowhammers the bits in memory,
they must also predictably corrupt the hash; otherwise, the attack will be
detected due to the hash mismatch. Thus, the attacker must profile the
effectiveness of Rowhammer on a cache line and not a word granularity, as they
have to be sure that the corrupted cache line and hash match. If the attacker
misses flipping one of the bits, the attack will fail with the probability of
$1-2^{-40}$ (i.e., the chance of collision for a 40-bit hash). In addition, the
attacker should consider that presence of even a simple ECC scheme increases the
time to rowhammer from minutes to days\cite{eccsploit}; with hashing, the time
to carry out the attack may increase to weeks, rendering this kind of
exploitation impractical. 

\subsection{Reliable Processing In-Memory}
The idea behind Processing In-Memory (PIM) is to compute near the data---i.e.,
in the main memory. Recently, PIM advanced from a theoretical research topic
toward a practical hardware implementation of the multiply-accumulate (MAC)
units within a commercial-grade HBM2 device\cite{lee2021hardwarepim}. While the
parity-based ECC seems to be a natural choice to protect the data of PIM
modules, \OURShort{} is more advantageous because it can protect both the stored
data and the computation itself without the need to convert the redundancy
information between the codes. For example, let us consider a setup used by Lee
\emph{et al.}~\cite{lee2021hardwarepim}: an HBM2 device with integrated MACs for
the acceleration of neural-network applications. In this device, the data are
read in 256b words and transferred to MACs for computation. To protect the data
and the compute, one can use ~\OURShort{}($268,\;256,\;\GM{}=3621$) code with
only 12 bits of redundancy. HBM provisions 64b for ECC storage for each 64B of
data, or 32b per 256b of data -- $2.6\times$ more than needed
for~\OURShort{}($268,\;256,\;\GM{}=3621$) code. The saved 20 bits from multiple
256b words provide enough space to store cryptographic authentication codes to
guarantee the integrity of the data. Moreover, PIMs may use error information to diagnose
the arithmetical units at runtime and guarantee reliable computation.

\section{Results}\label{sec:eval}
In this section, we study the following research questions: 
\begin{enumerate}
\item How do~\OURShort{} and Reed-Solomon code trade off storage for higher
reliability guarantees?
\item How do the VLSI overheads of~\OURShort{} compare to traditional codes?
\item How does the performance overhead of~\OURShort{} compare to Reed-Solomon
codes?
\item What are the benefits of co-designing~\OUR{} with memory tagging?
\item How does the flexibility of~\OURShort{} compare to a Reed-Solomon code?
\end{enumerate}

\subsection{Reliability Trade Off: Reed-Solomon
vs.~\OURShortnf{}}\label{sec:ck_codes} To compare Reed-Solomon and~\OUR{} codes,
we evaluate them across four parameters: (1) number of saved bits, (2)
practicality for DRAM, (3) single symbol error correction, and (4) multi-symbol
error detection rates. 

\fakesec{Saved Bits, Practicality, and Error Correction} 
Both Reed-Solomon and~\OURFull{} can be designed to offer space to store the
metadata. Reed-Solomon codes that offer saved bits, however, are not practical
in the context of DRAM memories because they may store more than one symbol in a
single chip. For example, let us assume a DIMM made with 4-bit DRAM devices
protected by Reed-Solomon code with 5-bit symbols. This code saves six bits of
storage, permitting 134 bits of data to be encoded into a 144-bit codeword. In
this scenario, the last bit of the first symbol and the first three bits of the
second are written to the same x4 device. If that device fails, the code will
not correct the error because there are two corrupted symbols instead of one.
Thus, this Reed-Solomon code cannot offer ChipKill error correcion. Hence, we
always need to evaluate bits savings in the context of the resulting code. 

\fakesec{Error Detection} 
\OURFull{} allows repurposing saved storage to gain higher multi-symbol error
detection rates. This can be done by choosing a larger multiplier value that can
be stored within the available redundancy bits. For example, we can design two
codes:~\OURShort{}(\MDDRoldMinSDC{}) and~\OURShort{}(\MDDRold{}). While both
codes offer single symbol correction, the first code gives up four saved bits to
store a larger multiplier value ($65519$ vs. $4065$) and gains higher
multi-symbol error detection:~\MDDRoldMinSDCMSED{}\% vs. 86.71\%. The
Reed-Solomon code, on the other hand, trades off both error correction and
detection guarantees to gain storage, effectively rendering those codes
impractical. However, while choosing zero saved bits, both RS(144,128)
and~\OURShort(\MDDRoldMinSDC{}) offer similar multi-symbol detection (MSED)
rates of 99.36\% and~\MDDRoldMinSDCMSED{}\%. If the saved bits are used store a
hash, as they are used in Rowhammer protection,~\OURShort{}'s MSED rate
increases even further to the chance of detectable hash collisions, which is
likely higher than the MSED rate achievable by a Reed-Solomon code, while
achieving better security.

Table~\ref{table:codes} shows MSED rates and bit-savings for various
Reed-Solomon and~\OURShort{} codes constrained to 144-bit codeword for x4
(5-bits savings shows~\OURShort{}(\MDDRnew{}) code). We see that both code
families provide spare bits at the cost of reduced multi-symbol error detection
rates. However, Reed-Solomon codes also lose ChipKill guarantees due to
misalignment of symbols with device boundaries because of the constant
two-symbol redundancy. For instance, to gain four spare bits with Reed-Solomon
codes, one must use 6-bit symbols (12-bit redundancy) to design RS(144,132)
code. This code is not practical because 6-bit-wide DRAMs do not exist. On the
other hand, with MUSE ECC, one simply picks a multiplier that guarantees
ChipKill correction and the desired bit savings. Moreover, due to their construction
constraints, Reed-Solomon codes offer bit savings only in multiples of two,
while MUSE allows for more fine-grained trade offs between the required storage
and MSED rates.

To compute multi-symbol detection rates, we used a Monte-Carlo-based simulator.
In an $n$-symbol codeword, the simulator randomly samples 10000 out of all ${n
\choose k}$ possible $k$-symbol error patterns. Each of the chosen $k$ symbols
is randomly corrupted, constructing a multi-symbol error. For each multi-symbol
error, we computed a syndrome and compared this syndrome to all the syndromes of
single symbol errors. If no match is found, then that specific multi-symbol
error is detectable. Additionally,~\OUR{} also detects multi-symbol errors
caused by the ripple of 1s or 0s across symbol boundaries. The fraction of
detectable multi-symbol errors, out of all sampled multi-symbol errors, is the
multi-symbol error detection rate. 

\begin{scriptsize}
\begin{table}[t]
 
 \caption{Design points of~\OUR{} and Reed-Solomon codes:\\
 Bit savings and Multi-Symbol Error Detection rates.}
 \centering

 \label{table:codes}
\begin{adjustbox}{max width=0.48\textwidth}
 \begin{tabular}{l | c | c | c| c| c| c| c}
 \toprule
 \multirow{2}{*}{\textbf{Code}} & \multicolumn{7}{c}{\textbf{Extra Bits}} \\
              & 0     & 1 & 2 & 3 & 4 & 5 & 6\\\hline
 RS           & \cellcolor{green!75} 99.36 & {\O} & \cellcolor{red!65}95.55 &
 {\O} & \cellcolor{red!65}86.79 & {\O} & \cellcolor{red!65}53.96\\\hline
 \OURShort{}  & \cellcolor{green!75}\MDDRoldMinSDCMSED{} &
 \cellcolor{green!75}98.35 & \cellcolor{green!75}96.70 &
 \cellcolor{green!75}93.39 & \cellcolor{green!75}\MDDRoldMSED{} &
 \cellcolor{blue!25}\MDDRnewMSED{} & {\O} \\
 \bottomrule
 \multicolumn{8}{l}{\textit{Legend:}}\\
 \multicolumn{1}{c}{{\O}} & \multicolumn{7}{l}{Code does not exist.}\\
 \multicolumn{1}{c}{\cellcolor{green!75}} & \multicolumn{3}{l}{144b ChipKill
 protection.} & \multicolumn{1}{c}{\cellcolor{blue!25}} & \multicolumn{3}{l}{80b
 ChipKill protection.}\\
 \multicolumn{1}{c}{\cellcolor{red!65}} & \multicolumn{7}{l}{Not practical code
 -- does not guarantee ChipKill protection.}\\
 \multicolumn{1}{c}{MSED} & \multicolumn{7}{l}{Error detection across multiple
 DRAMs at the same time.}\\

\end{tabular}
\end{adjustbox}
\end{table} 
\end{scriptsize}

\subsection{VLSI Overheads}

\begin{table*}
\small
 \centering
 \caption{Implementation results of the encoders and error correctors for \OURShort{}
 and Reed-Solomon ECC schemes.}
 \begin{adjustbox}{max width=\textwidth}
 \label{table:hw_impl}
 \begin{tabular}{r | r r r r | r r r r | r r }
 \toprule
 \multirow{2}{*}{\textbf{Code}} &\multicolumn{4}{c}{\textbf{Encoder}} &
 \multicolumn{4}{c}{\textbf{Error Corr. \& Det. }} &
 \multicolumn{2}{c}{\textbf{GEM5 Latency}} \\ 
  & \textbf{Latency}, \emph{ns} & \textbf{std\_cells} & \textbf{Area, $\mu m^2$}
  & \textbf{Power, $mW$} & \textbf{Latency}, \emph{ns} & \textbf{std\_cells} &
  \textbf{Area, $\mu m^2$} & \textbf{Power, $mW$} & Enc & Dec \\ \hline 
 \OURShort{}(\MDDRold{})    & 1.129 & 33312 & 10999 & 5.11 & 1.048 & 45493 &
 13648 & 8.56 & 3 & 0\\
 \OURShort{}(\MDDRnew{})	  & 1.177 & 11953 & 4166 & 5.22 & 1.179 & 18422 & 5593
 & 5.64 & 3 & 0\\ 
 \OURShort{}(\MDDRnewU{})	  & 1.154 & 14655 & 4896 & 4.14 & 1.018 & 24043 & 7092
 & 6.22 & 3 & 0\\
 \OURShort{}(\MDDRnewH{}) H & 1.181 & 13775 & 4772 & 4.15 & 0.859 & 18937 & 5719
 & 5.80 & 3 & 0\\
 RS(144,128)				& 0.219 & 1158 & 737 & 2.67 & 0.376 & 2884 & 1053 & 2.70 & 1
 & 0\\
 RS(80,64)				  & 0.124 & 542 & 359 & 1.31 & 0.381 & 2540 & 617 & 1.99 & 1 &
 0\\
 \bottomrule
 \end{tabular}
\end{adjustbox}
\end{table*} 

We implemented the basic arithmetic blocks, decoders, encoders and error
correctors in Verilog and synthesized them with Synopsys Design Compiler
Version: R-2020.09-SP4 using NangateOpenCell 15nm open-source standard cell
library\cite{os15nm}. The synthesis ran with hierarchy ungrouping and high
effort for delay, power, area optimizations. Table~\ref{table:hw_impl}
summarizes the latency and silicon area for the components of all discussed
codes.

The encoder latency of the~\OURShort{}(\MDDRold{}) code is 1.129\emph{ns}, while
the error correction latency (including remainder computation and ELC) is
1.048\emph{ns}. Assuming a clock frequency of 2400 MHz (or 417\emph{ps}), the
encoder delays the writes to the main memory by three clock cycles, while error
correction delays the reads by three clock cycles. For the systematic code, in
the common case of no errors, reads have no delay.

\fakesec{Reed-Solomon Code Implementation} We picked Reed-Solomon codes because they
are representative single-device correct ECC schemes. The Reed-Solomon
code is systematic; thus, no decoding is required. We picked the PGZ
algorithm\cite{wicker1995error} to implement the encoders and error correction
units for Reed-Solomon codes. In addition, for simplicity, we picked lookup
tables to implement Galois Field arithmetic.

Reed-Solomon encoders are simple XOR trees implementing binary multiplication of
generator matrix and data, resulting in low area overheads and single clock
cycle latency. However, error correction is more complicated as it requires
Galois Field arithmetic. Hence, the main factor differentiating
Reed-Solomon codes’ latency and silicon area of the error correction circuitry
is the symbol size, i.e., the number of entries in the lookup tables for symbol
arithmetic. The latency for error correction is 0.38\emph{ns}, area overheads
are between 842 to 1053$\mu m^2$, and power consumption ranges from 2 to
2.7\emph{mW} (See Table~\ref{table:hw_impl}). 

\fakesec{Comparison} Because the Reed-Solomon codes can be implemented with
simple XOR trees of moderate depth, they result in a smaller area and shorter
latencies than comparable~\OURShort{} codes. For example,~\OURShort(\MDDRnewU{})
code uses $12\times$ more silicon area than RS(80,64), adding two more clock
cycles of latency. These high area overheads are expected because Wallace Tree
nodes are two serially connected full adders, while XOR trees use a single XOR
gate. For the error correction, Reed-Solomon has a slightly smaller delay of
single a clock cycle vs. three cycles of~\OUR{}. The lead of Reed-Solomon codes
is not surprising, as those are natively suited for binary arithmetic. 

\subsection{Performance}
\begin{figure*}[t]
  \includegraphics[width=0.98\textwidth]{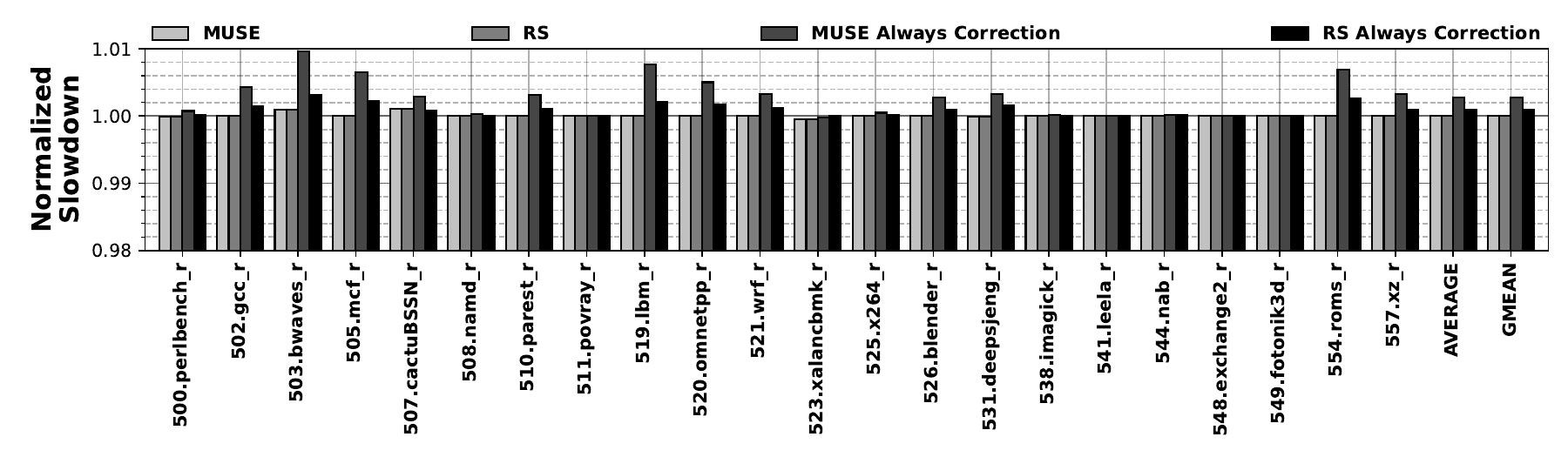}
  \caption{Normalized slowdown of a system due to the addition of ECC
  encoding/decoding on a memory interface. Modeled in gem5 using latencies from
  decoders/encoders synthesized with NangateOpenCell 15nm open-source
  process\cite{os15nm}.}
  \label{fig:latenciesSPEC}
\end{figure*}

\fakesec{Simulation Setup}\label{sec:performance} We evaluated the performance
impact of~\OURFull{} with the gem5 simulator
\cite{binkert2011gem5}
and SPEC 2017 (v1.1.5) benchmarking suite\cite{bucek2018spec}. We configured a
Haswell-like CPU with 3.4GHz frequency, 64kB L1 cache equally split for
instructions and data, L2 256kB/core, L3 of 8MB, and 32GB DDR4 memory. For the
simulation, we picked the TimingSimpleCPU model that provides a detailed timing
simulation of the memory subsystem while executing instructions in a single
clock cycle. We used gcc 4.8.5 to build dynamically-linked \texttt{fprate} and
\texttt{intrate} SPEC 2017 benchmarks with -O3 optimization level. We executed
the benchmarks with the reference inputs for 10 billion instructions, which is
long enough to warm up the caches and put the system in a steady state. 

To emulate encoding latency, we delay each write transaction on the memory
interface by the latency of the encoder. There is no decoding penalty
because all the evaluated codes are systematic. To estimate performance penalty
due to the delay of error correction, we delay memory reads by the latency of
the error-correcting circuit. To achieve this goal, we modified the memory
controller of the gem5 simulator. For convenience, the extra latencies are
summarized in the last two columns of Table~\ref{table:hw_impl}. 

\fakesec{Results and Discussion}
Figure~\ref{fig:latenciesSPEC} summarizes the slowdown of SPEC 2017 due to two
evaluated scenarios: (1) error-free operation, and (2) error correction on
every read operation. We see from the figure that~\OUR{} (blue bars) and
Reed-Solomon (orange bars) have similar performance to the baseline,
despite~\OURShort{} taking two more cycles for encoding than Reed-Solomon. This
is likely as write operations are rarely on a critical path of the system, and
delaying them by one or three cycles is unlikely to be critical for performance.
In the worst-case scenario where every memory read results in corrupted data,
Reed-Solomon (red bars) would have a slightly better performance
than~\OURShort{} (green bars)---a slowdown of 0.09\% vs. 0.2\% on average. As
we see from those results, performance overheads of~\OURShort{} are minimal and
comparable to those of the Reed-Solomon code.

\subsection{Case Study: Memory Tagging}\label{case:mte}

\begin{figure}[t]
\includegraphics[width=0.48\textwidth]{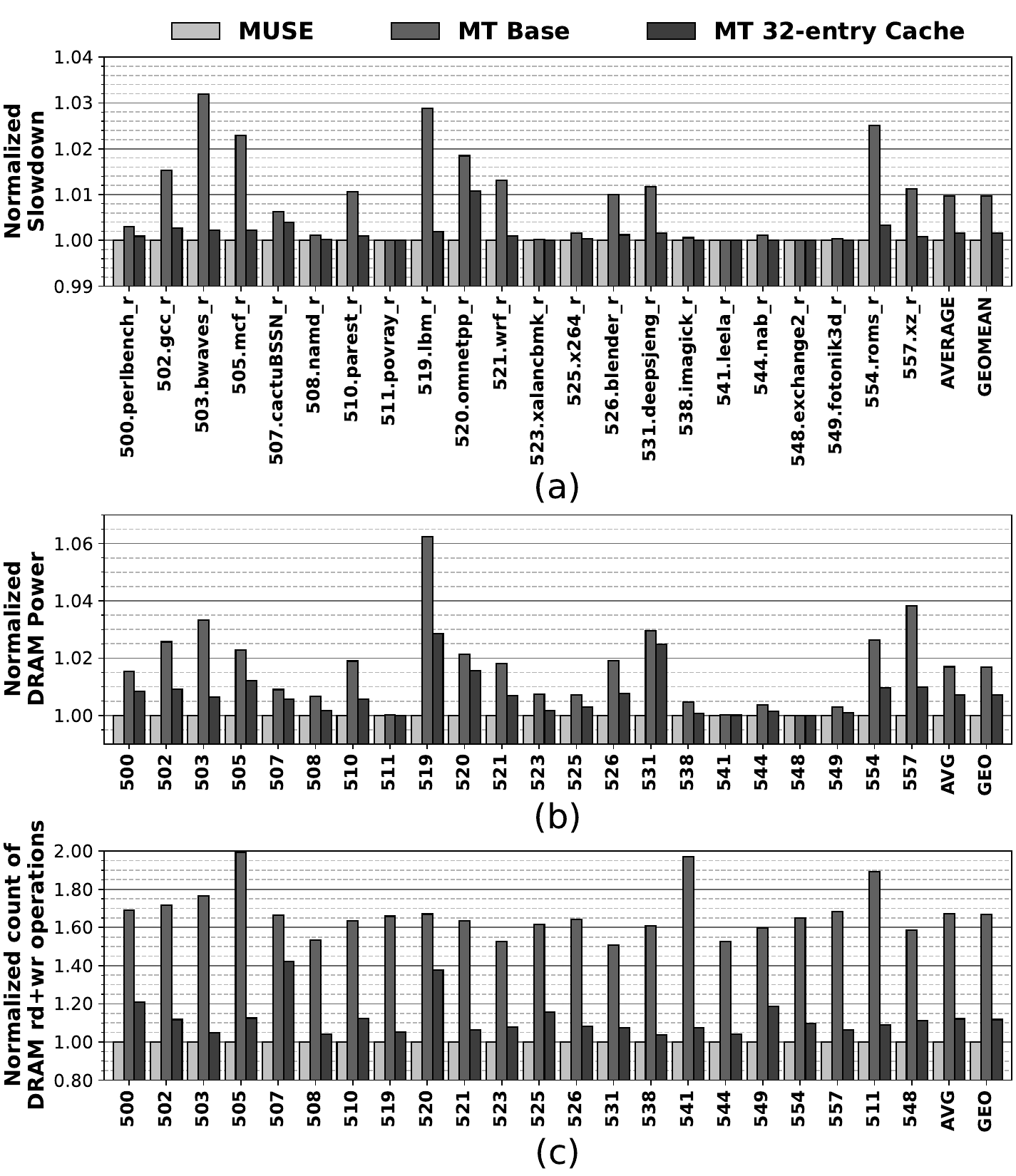}
\caption{Integration of Memory Tagging (MT) and~\OUR{}. All the results
  normalized to~\OURShort{}: (a) Normalized slowdown, (b) normalized DRAM power
  consumption, and (c) normalized number of read/write operations.}
\label{fig:mte_results_all}
\end{figure}

In this section, we analyze how the use of~\OUR{} with ARM MTE\cite{MTE}-like
memory tagging security scheme improves the power consumption of a system
without impact on performance. As a baseline, we assume a system with ECC and a
memory tagging scheme where every sixteen bytes of memory have a four-bit tag
(similar to ARM MTE), i.e., 16 bits of metadata for 512 bits of data. 

\fakesec{Overview} There are two approaches to implementing this scheme in
hardware. In the first approach, the metadata is stored in a disjoint manner in
a special region of the main memory. Thus, when the metadata is needed, an
additional memory request would bring it to the CPU from the main memory. To
mitigate the penalty of additional memory reads on performance, metadata caches
may be used to cache the metadata for later use, effectively reducing the number
of memory reads. The downsides of this approach are (1) more complicated design
(integration of metadata caches, additional state machines to track the
metadata state, etc.), and (2) increased power consumption due to metadata memory
traffic and metadata caches. The second approach is to store the metadata in the
ECC portion of the main memory, forgoing all ECC guarantees of a system. The
benefits of this approach are clear: (1) in-lined metadata, (2) no need for more
complex hardware, and (3) no increased power consumption as only the relevant
metadata is fetched from DRAM. The downside is that the system has no
ECC.~\OUR{} allows the system designer to enjoy the benefits of both approaches
by storing the in-lined metadata in the unused bits of the code to gain
performance, lower power consumption, simplify the design, and keep ChipKill-level ECC protection.

\fakesec{Evaluation Setup} 
We evaluate the following aspects of the system: (1) performance, (2) memory bus
traffic, and (3) power consumption. To do so, we modified the gem5 simulator to
issue an additional memory read for each cache miss to read a cache line worth
of metadata from reserved memory space. In addition, for the system with
metadata caching, we introduced a 32-entry 16kB metadata cache for memory tags.
We conservatively assume that metadata caches consume no power. We used the same
gem5 configuration as in Section~\ref{sec:performance} and evaluated the
following three configurations: (1) memory tagging with~\OUR{}, (2) Reed-Solomon
ECC with, and (3) without metadata caches. The SPEC-2017 benchmarks were run for
10B instructions on each of those configurations to measure execution latency,
DRAM power consumption, and the number of read/write transactions on the memory
bus.

\fakesec{Results and discussion} 
We summarize the simulation results in Figure~\ref{fig:mte_results_all}.
Figure~\ref{fig:mte_results_all}(a) shows that only the introduction of the
metadata caches can eliminate the overheads of memory tagging and equalize its
performance with that of MUSE---an improvement of about 1\% compared to
metadata-less memory tagging. Moreover, while the metadata caching reduces DRAM
power consumption from 1.7\% to 0.72\% (peaking at 2.8\% for 519.lbm), MUSE
still saves on average 0.72\% of DRAM power
(Figure~\ref{fig:mte_results_all}(b)). Finally,
Figure~\ref{fig:mte_results_all}(c) shows that while the metadata memory
requests improved with caching (67\% vs. 12\% on average), memory tagging with
caches still requires, on average, 12\% more metadata accesses than MUSE, which
requires zero additional memory requests due to its storage efficiency. These
metadata requests will result in the increased power consumption of the CPU's
memory controllers.

Table~\ref{table:powerb} summarizes the total power consumption of the evaluated
schemes. As we see from the summary, despite requiring more silicon area and
power for~\OUR{}, the overall system with memory tagging and~\OUR{} saves at
least $115$~\emph{mW} and 12\% of memory bandwidth while offering ChipKill
guarantees and a simpler system design.

\begin{scriptsize}
  \begin{table}[t]
   \centering
   \caption{Power consumption summary.}
   \label{table:powerb}
  \begin{adjustbox}{max width=0.48\textwidth}
   \begin{tabular}{lcccc}
   \toprule
   \multirow{2}{*}{\textbf{Scheme}} & \multicolumn{2}{c}{\textbf{Components,
   [mW]}} & \multicolumn{2}{c}{\textbf{}} \\
    & DRAM & ECC  & Total [mW] & diff, [mW]\\\hline
    MT w/~\OURShort{} & 6468 & $2\times14$ &  6496 & 0\\
    MT w/ 16kB cache & 6517 & $2\times5$ & 6527 & $+31$\\
    MT w/o cache & 6601 & $2\times5$ & 6611 & $+115$\\
  \bottomrule
   \end{tabular}
  \end{adjustbox}
  \end{table} 
\end{scriptsize}

\subsection{Flexibility: Reed-Solomon vs.~\OURShortnf{}}
\label{sec:design_rs_vs_muse} 
Here we evaluate the flexibility of Reed-Solomon and~\OURShort{} codes by
comparing how the code length, symbol size, and redundancy are interlinked and
their effect on the properties of a resulting code. We define the code as
flexible if its codeword and data length can be adjusted with fine resolution
(e.g., single-bit), the code can accommodate different symbol sizes  
multiple error models efficiently. 

For the Reed-Solomon code, redundancy, code, and data lengths are all functions
of the symbol size. For example, for a single-symbol error correcting
Reed-Solomon code, the redundancy is always two symbols long. Hence, there is
only one Reed-Solomon code for a given symbol size and codeword length.
Moreover, the Reed-Solomon code by design does not differentiate between bit
flip directions, making it a poor fit for the model of asymmetrical errors, as
the Reed-Solomon code will require the same number of redundancy bits for
asymmetrical and bidirectional errors. 

For~\OURFull{}, symbol size and code length define the set of valid code
multipliers---basically determining the number of redundancy and data bits in
the codeword. If the number of available data bits is more than needed, i.e.,
the code works with fewer redundancy bits, these saved bits can store the
metadata, be used for larger multiplier values improving error detection, or can
be discarded to create a shorter code (saving system resources). In
addition,~\OURFull{} allows the combination of multiple physical error models
into one code as long as a suitable multiplier is found.

\section{Related Work}\label{sec:rel}
\begin{figure*}[t]
    \includegraphics[width=\textwidth]{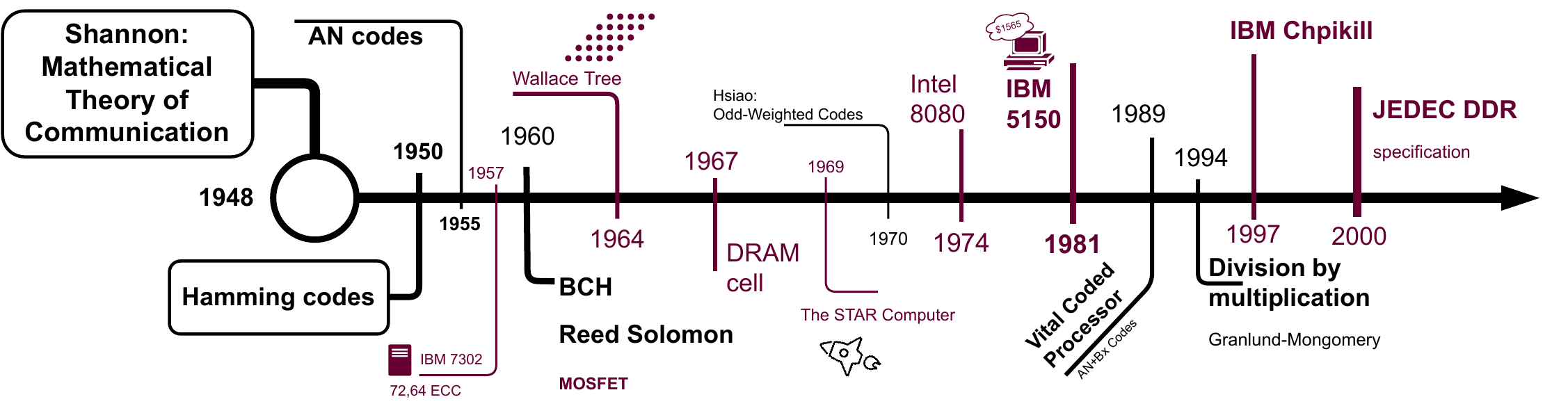}
    \caption{Historical timeline of significant developments with respect to residue
    codes.}
    \label{fig:timeline}
    \end{figure*}

\OURShort{} builds upon an interesting 60-year history of prior work on error correction 
(a brief historical overview is shown in
Figure~\ref{fig:timeline}).

Around the same time when BCH and Reed-Solomon codes were
developed\cite{hocquenghem1959codes,bose1960class,reed1960polynomial}, in 1960 Brown
published what is now referred to as an “AN” code\cite{brown_error_1960}.
The work was developed in the context of robust error correction for arithmetic
circuits. To generate a codeword for a data, a word $N$ it is multiplied by an
integer $A$. To retrieve the data word (and/or check for errors), the code word
is divided by $A$. In AN codes, if the sum of two codewords, say, $A \cdot{} x$
and $A \cdot{} y$, is also a code word, $A \cdot{} (x+y)$, then any error in
addition or subtraction can be identified by dividing the result by $A$ and
checking the error remainder. A variation of AN codes called residue codes
\cite{henderson_residue_1961,chien_linear_1964} were also developed in the
context of fault-tolerant arithmetic units. Instead of creating a code word as
$A \cdot{} N$, residue codes are created as $N \times P + A - B$, where $P$ is a
power of 2 and $B$ is the remainder when $N$ is divided by a multiplier $A$. The
nice thing about this construction is that it makes the code systematic, i.e.,
the data bits are separated from the check bits. This work and other
works\cite{mandelbaum_arithmetic_1967,barrows1966new} were published in the
pre-Moore’s-Law era, when there was a significant paucity of transistors
and when multiplications and divisions were too expensive to check
addition\cite{transistor_paucity,liu1970byte}. These ideas appear to have
remained of theoretical interest till the early 70s\cite{avizArith}. In 1970,
The STAR Computer\cite{STAR}---a research project funded by NASA to develop a
reliable computer for space probes\cite{TOPS}---used residue codes. Since then,
residue codes continue to be used to guarantee the reliability of arithmetic
units in the CPU (e.g., \cite{ibm_resid}). In 2010, Sullivan~\cite{msc_sul}
systematically evaluated the overheads of residue checking for various error
models in the context of modern ALUs.

In 1967, Dennard invented DRAM\cite{dram_patent}, and soon after there was a
flurry of work on improving the reliability of these new ``high-speed''
memories\cite{hsaio_review}. While technology scaling in the 1970s permitted the
creation of single-chip microprocessors in this era, latency was still a
significant concern, and it was believed that high-speed memories would need
error correction techniques that can ``encode and decode'' in
parallel\cite{rao1989error}. Alas, the arithmetic and residue error codes cannot
be parallelized unless significant restrictions are placed on the multiplier. In
1970, Hsiao published his seminal work on Odd-parity weight
codes\cite{hsiaoCode} that balanced the delay of the syndrome calculation for a
SEC-DED code using only seven levels of fast XOR gates. Thus, the idea of AN and
residue codes was put aside again.

In the 1990s, with companies like Rambus, DRAM became even faster due to
innovative signaling techniques, which cemented the place for lightweight
Hsiao-like codes; since then, these constructions have remained the de facto
standard for many systems. Effectively by this time, the value of AN and residue
codes appear to have disappeared from the collective memory of the research
community as evidenced by the lack of papers on this topic. At the same time, in
a completely different community, a technique that is key to unlocking the
potential of these codes was being developed.

In 1994, Granlund and Montgomery published a paper ``Division by invariant
integers using multiplication''\cite{gccDivByConst} at the PLDI Conference. They
observed that computers were multiplying much faster than dividing---around one
order of magnitude---so they asked the question: would it be possible to perform
a division as a multiplication? For division by constants, this is clearly
possible, especially if the constants can be pre-computed at the compile time of
the program. For instance, to divide a number by $5$, we can multiply it by
$0.2$. To avoid the floating-point multiplication, we first multiply the inverse
of $5$ by a large integer power-of-two and then divide the result by the same
integer power-of-two. We replace the last division with a very cheap shift
operation. We leverage this advance to make \OURShort{} hardware very fast.

In the early 90s, most server systems still employed parity for error detection.
The race for more reliability and availability features increased with the rise
of e-commerce websites in the late 90s. IBM introduced “ChipKill”
memory\cite{chipkillIBM,enhancing1999netfinity}, a commercial name for a system
that continues operation even if an entire DRAM device fails. Basically, in the
late 90s, the error model moved from single errors to supporting burst errors.
These availability features were also used in space missions, such as NASA’s
pathfinder MARS probe\cite{enhancing1999netfinity}. In contrast to these popular
burst codes, \OURShort{} provides the same reliability with fewer check bits.

In 2000, JEDEC standardized the DDR interface to memory\cite{jedec2000jesd79}.
Standardization meant interoperability, but it also introduced several
constraints on how memory would be integrated, and, pertinently, how error
correction could/should be performed. DDR4 memory ECC DIMMs are 72-bit wide:
64-bit data and 8-bit check bits. This architecture can be used to implement a
(72,64) code, or a (144,128) code\cite{devicesbios}. Typically, the (72,64) code
supports SEC-DED using Hsiao or Hamming codes, and (144,128) is used for single
and burst error correction. With the continuing technology and voltage scaling
trends, DRAM has suffered more and more reliability problems, especially
Low-Power DDR memories (used in mobile phones and embedded devices). As
mitigation, vendors started to implement on-die ECC\cite{oh20143,isslpddr4ecc},
typically as simple single error correcting codes without the ability to correct
burst-errors. The idea of using on-die error correction has now been
standardized in the latest generation of JEDEC’s DDR5
standards\cite{jedec2020jesd796}.

For the modern servers and mainframes, the reliability requirements are more
demanding. For example, AMD\cite{amdEpyc2017,devicesbios}, IBM\cite{power9ras},
and Intel\cite{supermicro,dddc} systems offer ChipKill-level guarantees for
DIMMs with x4 and x8 devices. Due to ECC construction constraints, standard x8
DIMMs do not have enough DRAM chips to offer ChipKill over a single 72b channel.
To mitigate this, IBM manufactures custom DIMMs with x8 DRAMs, uses 144-bit
busses between the memory controller and the DIMMs, and makes cache lines
128-byte long in the POWER9 series CPUs\cite{power9ibm,power9ras}. With these
techniques in place, IBM systems support at least a single device failure. For
x4-based DIMMs and custom x8 DIMMs, some IBM systems may tolerate two
consecutive device failures on a pair of DIMMs\cite{power9ras}. Similar
guarantees are provided by Intel for x4-based DIMMs with its Double Device Data
Correction (DDDC) scheme\cite{dddc,supermicro}. However, because Intel systems
are mostly based on off-the-shelf components, DDDC does not support x8 DIMMs.
While~\OUR{}'s support for x8 DIMMs is to be developed, we offer protection
against two consecutive device failures for x4 DIMMs, and, unlike commercial
solutions, our code has the ability to be tuned for a specific error model.  

Academic ECC schemes tend to be more diverse than current commercial solutions.
These proposals range from a new codeword organizations
(e.g.,~\cite{kim_bamboo_2015,lowpolowsto}), ECC enablement in systems without
ECC (e.g., hashing and checkpointing\cite{memguard}, virtualization of
ECC\cite{yoon2010virtualized}, memory compression for ECC\cite{kim2015frugal}),
extension of ECC to both DRAM data and control signals\cite{kim2016all}, use of
small caches within the DRAM\cite{CIDRA,care_ecc} to multi-tier and/or
concatenated codes (e.g., \cite{cleanecc,lotEcc,yoon_memory_2009}).  
More recent academic ECCs design considering the on-die ECC of modern DRAMs. For
example, XED\cite{XED} and PAIR\cite{pairecc} work in conjunction with on-die
ECC, while DUO\cite{duoEcc} repurposes on-die ECC's storage to increase the
number of bits for redundancy of the memory controller's ECC, thus increasing
the strength of the resulting scheme. Those, and many other solutions, consider
ECC only for its main purpose---to achieve reliable, error-free system
operation. \OURShort{}, on the other hand, optimizes for two objectives at the
same time: ChipKill-level error correction \textit{and} availability of in-lined
metadata. Nonetheless, on-die ECC is an integral part of the new DDR5
DRAM\cite{jedec2020jesd796}, and, thus, we believe that the investigation of
\OURShort{} co-design with on-die ECC is an interesting topic for future work.

However, ECC may improve not only the reliability but also the power consumption
of a system. For instance,~\cite{cahce_life_ecc,miller2010parichute} use strong
ECC schemes to enable reliable operation of caches in near-threshold regime,
saving 34 to 71\% of system power. However, these power savings come at the cost
of trading off a significant portion of the cache to store ECC bits (up to
50\%), which may be an acceptable trade-off for a system in a low-power mode. In
contrast, we designed \OURShort{} to minimize the memory footprint of the ECC to
improve system performance, power efficiency, and memory bandwidth utilization.
Due to \OURShort{}'s storage efficiency, it may be a perfect candidate for
caches operating in near-threshold regimes, but a detailed cost-benefit analysis
requires a detailed study which we leave for future work.

Along with the growth in reliability problems, the last decade also witnessed an
increase in security problems. Several recent security solutions increase the
demand for memory, including memory encryption\cite{SEV}, adding authentication
code\cite{costan2016intel,liljestrand2019pac}, or requiring additional storage
to support software security (e.g.,~\cite{MTE,zero_pointer,Califorms,cheri256}).
We developed \OUR{} to support both error correction and security. Prior works
in the area of co-designing security and reliability assume standard ChipKill
or SEC-DED codes\cite{huang2010ivec,isca1983, irea_erea, taassoricompact,
saileshwar2018synergy}, and those codes do not provide a way to extract more
states for security or performance features or reduce the amount of storage for
reliability which are key contributions of \OUR{}.

\section{Conclusion}\label{sec:con}
In this paper, we present~\OUR{}---a novel ECC construction that provides a
ChipKill-level of error correction but crucially also allows spare ECC space to
be harvested and used for additional purposes. To make~\OUR{} work in modern
settings, we revisit old formulations of residue codes and extend them with two
novel optimizations---symbol errors and shuffling of bits. With these
optimizations, we show that~\OUR{} can act as a ``drop-in'' replacement for
commonly used ChipKill ECC schemes while using fewer bits for storage. Moreover,
when we evaluated~\OUR{} ECC in ensemble with security technique of Memory
Tagging, we found out that the system is more power efficient than system with
conventional Reed-Solomon codes. The benefits of~\OUR{} codes also go beyond
space savings and holding metadata. For instance,~\OUR{} codes integrate much
more easily into Processing In-Memory devices, making~\OUR{} a very exciting ECC
option for the next decade and beyond.

\section*{Acknowledgment}\label{sec:ack} This work was partially supported by
Qualcomm Innovation Fellowship. Any opinions, findings, conclusions and
recommendations expressed in this material are those of the authors. Simha
Sethumadhavan has a significant financial interest in Chip Scan Inc.

\appendix
\section{Artifact Appendix}\label{sec:ae}

\subsection{Abstract}
This artifact contains the following three components
\begin{enumerate}
\item \textbf{Multiplier Search}: the implementation of the code search
procedure outlined in Algorithm 1 in the paper. 
\item \textbf{gem5 simulator}: modified gem5 simulator for performance analysis
reported in Figures~\ref{fig:latenciesSPEC} and~\ref{fig:mte_results_all}). 
\item \textbf{VLSI implementation}: Verilog implementation of~\OUR{} and
Reed-Solomon codes used in the paper.
\end{enumerate}
Artifacts (1) and (2) are provided in the form of the Docker container, so they
are easy to set up and run. Artifact (3), however, has only the Verilog code as
we used proprietary tools we cannot share. However, (3) is easy to reproduce
once you have access to the tools as we used a default Synopsys RM-Flow (see
Section~\ref{ap:verilog}). Each of the provided artifacts has a detailed
README.md file with instructions. 

\subsection{Artifact check-list (meta-information)}
{\small
\begin{itemize}
  \item {\bf Algorithm:} Code search procedure (Algorithm 1 in the paper) 
  \item {\bf Compilation:} performed automatically within the docker container.
  \item {\bf Run-time environment:} Docker container.
  \item {\bf Hardware:} x86\_64-based system.
  \item {\bf Execution:} No requirements. 
  \item {\bf Metrics:} Valid code multipliers are reported if found, figure pdf
  files. 
  \item {\bf Output:} A text file per experiment with search configuration
  supplied via command line, i.e., codeword and symbol lengths in bits,
  redundancy budget in bits, etc., and a list of found multipliers. Both figures
  as pdf files.
  \item {\bf Experiments:} Four experiments are required to reproduce the codes
  reported in Table 1. In addition, the user may change command line arguments
  to see how those affect code feasibility, i.e., the configuration of
  ~\OUR{}(80,67) code without \emph{shuffling} finds no multipliers. 280 gem5
  simulations. 
  \item {\bf How much disk space is required (approximately)?:} about 30 GB. 
  \item {\bf How much time is needed to prepare workflow (approximately)?:}
  10-20 minutes to build the containers (depending on the internet connection).
  \item {\bf How much time is needed to complete experiments (approximately)?:}
  Code Search: less than 20 minutes to build the container and run all the
  experiments (automated via docker image). gem5 simulations require 36 hours on
  M5zn AWS instance (48 cores). 
  \item {\bf Publicly available?:} Yes.
  \item {\bf Code licenses (if publicly available)?:} APACHE 2.0
  \item {\bf Workflow framework used?:} Docker containers.
  \item {\bf Archived (provide DOI)?:} \url{https://doi.org/10.5281/zenodo.7019209}
\end{itemize}
}

\subsection{Description}

\subsubsection{How to access}
The artifact is licensed with Apache 2.0 license and can be downloaded from
\url{https://doi.org/10.5281/zenodo.7019209}. The artifact contains the source code,
installation instructions, and steps to run the experiments. 

\subsubsection{Hardware dependencies}
The artifact was developed and tested on x86\_64-based system with Intel Core
i7-8700 CPU.

\subsubsection{Software dependencies}
For the code search, the artifact requires boost \texttt{C++} library v1.74 and
\texttt{g++} (\texttt{v10} or \texttt{v11}) compiler. For the gem5 simulations,
SPEC 2017 benchmarking suite is required. For Verilog synthesis, Synopsys Design
Compiler, PrimeTime PX tools, and NangateOpenCell 15nm open-source standard cell
library are required (those are not included in the artifact, the library can be
obtained for free via \url{https://si2.org/open-cell-library}). 

\subsection{Installation}
\fakesec{Multiplier Search.} The code search supports two ways of installation:
(1) automated via docker container or (2) manual compilation. 
\begin{itemize}
\item The preferred way to evaluate the artifact is to use provided docker
container: download and install docker daemon via
\url{https://docs.docker.com/get-docker/}. Build the container in the root
folder of the artifact with \hlc[gray!25]{\texttt{docker build -t muse:muse~.}}.
The docker image both builds the source code and performs the multiplier search. 
\item Alternatively, install boost library via \hlc[gray!25]{\texttt{sudo apt
install libboost-all-dev}} and compile the source code with
\hlc[gray!25]{\texttt{g++ --std=c++17 -pthread code\_search.cpp}}.
\end{itemize}
For convenience, the artifact comes with \texttt{README.md} file explaining both
of these options. 

\fakesec{Simulations with gem5.} 
gem5 simulations are performed in the provided docker container which includes a
modified version of the gem5 simulator, which can be built with
\hlc[gray!25]{\texttt{docker build -t muse-gem5:muse-gem5~.}} User is required
to have their copy of the SPEC 2017 benchmarking suite installed at
\hlc[gray!25]{\texttt{/all\_data/spec2017}}. Note, that SPEC 2017 installation
location is customizable, see included README.md for more details.   

\subsection{Experiment workflow}
\fakesec{Multiplier Search.} The experiments are performed automatically during
the container build process. Multiplier search results can be examined by
opening the container interactively: \hlc[gray!25]{\texttt{docker run -i -t
muse:muse bash}}.Once the shell is available, four text files with the
\texttt{.result} extension contain the set of found code
multipliers for each executed search configuration.

To execute the experiments manually, run the compiled binary with configuration
matching desired code constraints. For example, to run the search for 144-bit
codewords with a 12-bit redundancy budget (i.e., ~\OUR{}(144,132) code), use the
following command line parameters: 

\begin{itemize}
  \item 12-bit multipliers \hlc[gray!25]{\texttt{-p 12}}
  \item 144-bit codewords \hlc[gray!25]{\texttt{-b 144}}
  \item 4-bit symbols \hlc[gray!25]{\texttt{-m 4}}
  \item (optional) multithreading \hlc[gray!25]{\texttt{-t 8}}
\end{itemize}
Please refer to the \hlc[gray!25]{\texttt{--help}} output for the remaining
command line options of the compiled binary or see the examples in
\texttt{README.md}.

\fakesec{Simulations with gem5.} The simulations should be launched via
\hlc[gray!25]{\texttt{make sim}}. When the simulations are finished, the figures
are plotted with \hlc[gray!25]{\texttt{make figures}}. As a result, Figure6.pdf
and Figure7.pdf will appear. Please refer to included README.md for more details
about the manual line-by-line process. 

\subsection{Evaluation and expected results}
\fakesec{Multiplier Search.} Four experiments are performed automatically during
the installation steps of the artifact for each reported \OUR{} code in Table 1
in the paper. Each search configuration will result in at least one found
multiplier (\hlc[gray!25]{\texttt{highlighted values}} are used for the codes
reported in the paper): 

\begin{itemize}
  \item For 144b codewords with 12-bit redundancy, and 4-bit symbols: 2397,
  2883, 2967, 3009, 3259, 3295, 3371, 3417, 3431, 3459, 3469, 3505, 3523, 3531,
  3551, 3555, 3621, 3679, 3739, 3857, 3909, 3995, 4017, 4043,
  \hlc[gray!25]{\texttt{4065}}.
  \item For 80b codewords with 11-bit redundancy, and 4-bit symbols: 1491, 1721,
  1763, 1833, 1875, 1899, 1955, \hlc[gray!25]{\texttt{2005}}.
  \item For 80b codewords with 13-bit redundancy, asymmetrical symbol errors,
  shuffling, and 8-bit symbols: \hlc[gray!25]{\texttt{5621}}.
  \item For 80b codewords with 10-bit redundancy, asymmetrical symbol errors and
  all single-bit errors, shuffling, and 4-bit symbols:
  \hlc[gray!25]{\texttt{821}}.
\end{itemize}
These results are summarized in \texttt{README.md} as well. 

\fakesec{Simulations with gem5.} At the end of the simulations,
Figure~\ref{fig:latenciesSPEC} and Figure~\ref{fig:mte_results_all} will be
generated, and should match those from the paper. 

\subsection{Experiment customization}
To see the impact of shuffling on the code search, one may perform the search
for \OUR{}(80,70) code without shuffling, i.e., by specifying
\hlc[gray!25]{\texttt{-s 0}}. In this case, no code multiplier would be found.

\subsection{Verilog Synthesis and Analysis}\label{ap:verilog}
Here we outline a brief guide to synthesize and analyze the designs.
\begin{enumerate}
\item Request OpenCell NanGate 15nm standard cell library here (free): https://si2.org/open-cell-library/.
\item Download synthesis RM-Flow for the Design Compiler from
\url{https://solvnet.synopsys.com/rmgen/} by selecting ``Design Compiler'' in
the dropdown menu (free with Synopsys license).
\item Modify your synthesis script to include those lines instead of generic \hlc[gray!25]{\texttt{compile}} command:
\begin{itemize}
  \item \hlc[gray!25]{\texttt{ungroup -flatten -all}}, and
  \item \hlc[gray!25]{\texttt{compile -boundary\_optimization -map\_effort high -area\_effort high -power\_effort high -auto\_ungroup delay}}
\end{itemize}
\end{enumerate}

For the power analysis with Synopsys PrimeTime PX use the VCD-driven
RM-Flow from \url{https://solvnet.synopsys.com/rmgen/} (pick ``PrimeTime'' in
the dropdown menu).

\subsection{Notes}
\fakesec{Multiplier Search.}
We use the boost library for its long integers (i.e., 256-b, etc.). Since this
feature is not unique to v1.74, we suppose older versions may work too.

\fakesec{Simulations with gem5.} Provided docker container may be slightly
different from the environment we used on the university cluster; thus, minor
differences with the paper results may be observed. 

\fakesec{Verilog.} We cannot share proprietary tools we used for the synthesis
and power analysis. Please refer to the included README.md for directions to
reproduce the results with your own Synopsys tools.

\subsection{Methodology}

Submission, reviewing and badging methodology:

\begin{itemize}
  \item \url{https://www.acm.org/publications/policies/artifact-review-badging}
  \item \url{http://cTuning.org/ae/submission-20201122.html}
  \item \url{http://cTuning.org/ae/reviewing-20201122.html}
\end{itemize}

\bibliographystyle{IEEEtranS}
\bibliography{main}
\end{document}